\documentclass[aps,pra,twocolumn,10pt]{revtex4-1} 
\usepackage{amsmath,amsfonts,amssymb,amsthm}
\usepackage{mathrsfs}
\usepackage{graphicx,rotating,booktabs}
\usepackage{epsfig} 
\usepackage{setspace}
\usepackage{array}
\usepackage{braket} 
\usepackage{bbm} 
\usepackage{xcolor} 
\usepackage{verbatim}
\usepackage{siunitx} 
\usepackage{times} 
\usepackage{mathtools} 
\usepackage{mathrsfs} 
\RequirePackage{color} 
\RequirePackage[colorlinks=true, urlcolor=blue, anchorcolor=blue, citecolor=blue, filecolor=blue, linkcolor=blue, menucolor=blue, linktocpage=true, pdfproducer=medialab, pdfa=true]{hyperref} 
\DeclareFontFamily{OT1}{pzc}{}
\DeclareFontShape{OT1}{pzc}{m}{it}{<-> s * [1.10] pzcmi7t}{}
\DeclareMathAlphabet{\mathpzc}{OT1}{pzc}{m}{it}
\DeclareFontFamily{U}{BOONDOX-calo}{\skewchar\font=45 }
\DeclareFontShape{U}{BOONDOX-calo}{m}{n}{
 <-> s*[1.05] BOONDOX-r-calo}{}
\DeclareFontShape{U}{BOONDOX-calo}{b}{n}{
 <-> s*[1.05] BOONDOX-b-calo}{}
\DeclareMathAlphabet{\mathcalboondox}{U}{BOONDOX-calo}{m}{n}
\SetMathAlphabet{\mathcalboondox}{bold}{U}{BOONDOX-calo}{b}{n}
\DeclareMathAlphabet{\mathbcalboondox}{U}{BOONDOX-calo}{b}{n}

\newcommand{\eqd}{\vcentcolon=}

\newcommand{\abs}[1]{\left\lvert#1\right\rvert}
\DeclareMathOperator{\tr}{tr}

\newtheorem{prop}{PROPOSITION}

\newcommand{\be}{\begin{equation}}
\newcommand{\ee}{\end{equation}}
\begin{document} 
\title{Trade-off between information and disturbance in qubit thermometry} 
\author{Luigi Seveso}
\email{luigi.seveso@unimi.it} 
\affiliation{Quantum Technology Lab, Dipartimento di Fisica dell'Universit\`a degli Studi di Milano, I-20133 Milano, Italia} 
\author{Matteo G. A. Paris}
\email{matteo.paris@fisica.unimi.it} 
\affiliation{Quantum Technology Lab, Dipartimento di Fisica dell'Universit\`a degli Studi di Milano, I-20133 Milano, Italia \\
Istituto Nazionale di Fisica Nucleare, Sezione di Milano, I-20133 Milano, Italy\\ 
Department of Mathematics, Graduate School of Science, Osaka University, Toyonaka, Osaka 560-0043, Japan}
\begin{abstract} 
We address the trade-off between information and disturbance in qubit thermometry from the perspective of quantum estimation theory. Given a quantum measurement, we quantify information via the Fisher information of the measurement and disturbance via four different figures of merit, which capture different aspects (statistical, thermodynamical, geometrical) of the trade-off. For each disturbance measure, the {\em efficient} measurements, i.e.~the measurements that introduce a disturbance not greater than any other measurement extracting the same amount of information, are determined explicitly. The family of efficient measurements varies with the choice of the disturbance measure. On the other hand, commutativity between the elements of the probability operator-valued measure (POVM) and the equilibrium state of the thermometer is a necessary condition for efficiency with respect to any figure of disturbance.  
\end{abstract}  
\pacs{}   \keywords{} 
\maketitle 
\section{Introduction}\label{s:intro} 
Extracting information from a physical system by performing a quantum measurement disturbs the original state of the system. This fact has been known since the early days of quantum mechanics, e.g.~it lies at the basis of the original formulation of the uncertainty principle \cite{heisenberg1985anschaulichen}. More recently, it has been recognized not only as a limiting factor, but also as a resource for various quantum information processing tasks, such as quantum cryptography \cite{bennett1992quantum,ekert1991quantum,bennett2014quantum}. Intuitively, the greater the amount of information extracted, the greater the disturbance caused by the measurement. This intuition has found several quantitative expressions over the years, which differ by the choice of how to quantify the information and the disturbance associated to any given measurement scheme \cite{fuchs1996quantum,banaszek2001fidelity,fuchs2001information,d2003heisenberg,maccone2006information,cvtoff,ozawa2004uncertainty,genoni2005optimal}. Most studies have focused on the trade-off relation between information and disturbance in a purely information-theoretic setting, e.g.~the measurement extracts information about a message encoded in a quantum state and the disturbance is quantified via a fidelity-based distance between the original and the post-measurement state. Here, for a  variety of reasons outlined below, we focus on a different framework. 
\par
Our analysis is set in the context of {the theory of quantum parameter estimation} \cite{pla95,holevo2011probabilistic,h72,braunstein1994statistical,paris09}. The typical quantum parameter estimation task is, given  {a one-parameter family of quantum states $\rho_\xi$ (referred to as a  quantum statistical model),} to infer the true value of $\xi$  {via repeated measurements on $\rho_\xi$ and a suitable post-processing of the outcomes}. The precision achievable by any estimation strategy is expected to be inversely related to the disturbance it caused. We focus in particular on the case where the unknown parameter is the temperature of a thermal bath, i.e.~quantum thermometry, since it provides a natural testbed for the exploration of the information/disturbance trade-off from the estimation perspective.  {Indeed, upon performing a measurement on a state at thermal equilibrium, to which a temperature can be meaningfully assigned, one generally obtains an out-of-equilibrium state and the question naturally arises of how disturbance should be quantified.}
\par
In recent years, interest has been growing in the use of individual quantum systems for temperature estimation \cite{campbell2017global,jevtic2015single,campbell2018precision,hofer2017quantum,correa2015individual,luis17,hofer2017fundamental}. Micro-mechanical resonators received much attention \cite{nmr1,nmr2,nmr3,nmr4,nmr5,nmr6,nmr7,nmr8,nmr9,nmr10,brunelli2011qubit,brunelli2012qubit}, but also other viable platforms have been proposed, from quantum dots \cite{walker2003quantum} to SQUIDs \cite{halbertal2016nanoscale}, and NV centers in diamond \cite{toyli2013fluorescence,neumann2013high,kucsko2013nanometre}. The necessity to study the interplay between quantum mechanics and thermometry is due to the fact that increasing spatial resolution requires the probe to be so small that eventually quantum-mechanical behavior becomes inescapable. Moreover, quantum effects, such as coherence and entanglement, promise to be useful resources in their own right to enhance sensitivity beyond what is classically achievable. 
\par
The typical thermometry protocol involves bringing the probe in contact with the sample, treated as a thermal bath, and  {wait long enough for it} to thermalize. Information about the temperature is thus encoded into the  {equilibrium} state of the probe. The statistics of the outcomes,  {for a suitable measurement performed on the probe, allows one} to infer the temperature of the bath. In a standard estimation scenario, one is interested in extracting as much information as it is allowed by quantum mechanics. The optimal measurement to perform is then a projective measurement of the symmetric logarithmic derivative of the statistical model, i.e., in the context of thermometry, an energy measurement. However, such a measurement is highly disturbing, since the post-measurement state carries no residual information about the temperature. If one needs to monitor the temperature in time, one would prefer to implement a measurement which, while extracting a non-vanishing amount of information, does not disturb the system so dramatically. For  {a given value of the extracted information, an \emph{efficient} measurement is that introducing a disturbance not greater than any other measurement.} Efficient measurements make up the frontier of the trade-off region in the plane information vs disturbance. 
\par
In this  {paper}, we investigate the trade-off between the information on the temperature of a thermal bath extracted via a  {quantum} measurement on a probe, which plays the role of thermometer, and the disturbance that the thermometer itself suffers as a result. In particular, we employ an individual qubit as a probe and quantify the information in terms of the Fisher information, which has a direct statistical interpretation \cite{rao1992information,cramer2016mathematical}. The choice of the disturbance measure is, however, less clear-cut. In the following, four different disturbance measures are  {put forward, defined and evaluated}. We quantify the disturbance via the information-loss \cite{shitara2016trade}, the fidelity-based distance between the original and post-measurement state \cite{fuchs1999cryptographic} and two other disturbance measures: the first has a quantum thermodynamical origin \cite{gemmer2004quantum} and the second an information-geometrical interpretation \cite{amari2007methods}. Precise definition are given in the next section. In particular, we study how each of these measures is correlated to the POVM \emph{purity} $\gamma$ and \emph{non-commutativity} $\chi$ (to be  {explicitly} defined in the following). We investigate their trade-off with the extracted information and determine, either analytically or numerically, the efficient measurements. While the resulting trade-off regions vary depending on the chosen disturbance measure, some general features emerge: {\bf 1.} for fixed value of the non-commutativity, the measurements that maximize the disturbance are the projectives ones; {\bf 2.} for fixed value of the purity, the measurements that maximize the disturbance are the irreversible ones; {\bf 3.} the efficient measurements always belong to the family of semiclassical measurements. Semiclassical measurements   {are those minimizing} the non-commutativity, i.e.~they commute with the thermal state of the probe before the measurement.  The fact that all four disturbance measures lead to comparable results suggests, on the one hand, that all four  measures, though capturing different aspects of the trade-off relation,  {are meaningful in their own.} On the other hand, they suggest that such features could apply more generally,  {beyond} the specific model adopted here. 
\par
The rest of the paper is organized as follows. In Sec.~\ref{s1}, we introduce the necessary definitions and specify how information and disturbance are to be quantified. In Sec.~\ref{infoqt}  {we introduce the  {set of all measurements} to be considered, and analyze in details how the extracted information varies} as a function of the POVM parameters. In Sec.~\ref{distqt}, a similar analysis is carried out for the four different measures used to quantify disturbance, with particular attention to the trade-off relation with the extracted information. In Sec.~\ref{concl}, we summarize the main results and draw {a few general} conclusions from our work. 
\section{Preliminaries}\label{s1}
\subsection{Notational conventions}
We restrict ourselves to the case of a finite-dimensional quantum system with Hilbert space $\mathcal H = \mathbbm C^d$. The space of  $d\times d$ complex (\emph{resp.}, Hermitian, Hermitian positive semidefinite) matrices is denoted by $\mathsf M_d$ (\emph{resp.}, $\mathsf{Her}_d$, $\mathsf{Her}^+_d$). $\mathbbm I_{\mathcal H}$ stands for the identity matrix on $\mathcal H$. 
\par
A quantum statistical model is a family of density operators $\rho_\xi \in \mathsf{Her}^+_d$, parametrized by a real parameter $\xi \in \Xi \subset \mathbbm R$, where $\Xi$ is referred to as the parameter space. The parametrization map $\xi \to \rho_\xi$ is injective and as smooth as required. The unknown parameter $\xi$ is alternatively denoted by $\beta$ in the special case when it is the inverse temperature of a thermal quantum statistical model $\rho_\beta = e^{-\beta H}/Z_\beta$, with $H \in \mathsf{Her}_d$ being the system's Hamiltonian and $Z_\beta = \tr(e^{-\beta H})$ its partition function. 
\subsection{ {Quantifying information}}
We fix our notation concerning quantum measurements and then introduce the Fisher information of a measurement,  {which quantifies the maximum information that a measurement extracts about a parameter.}
\par
Assuming for ease of notation that the sample space $\mathcal X$ of the measurement is discrete, a measurement scheme $\mathscr M$ is defined in terms of its corresponding positive-operator valued measure (POVM) $\{\Pi_x\}_{x\in \mathcal X}$, with $\Pi_x \in \mathsf{Her}^+_d$ and $\sum_{x\in \mathcal X} \Pi_x=\mathbbm I_{\mathcal H}$. The probability $p_{x,\,\xi}$ of a given outcome $x\in\mathcal X$ is given by $p_{x,\,\xi}=\tr(\rho_\xi \Pi_x)$.
\par
A measurement scheme, while specifying the statistics of the observed outcomes, does not specify how the state of the system is updated as a result of the measurement. An \emph{instrument} $\mathcal I$, corresponding to the measurement scheme $\mathscr M$, is a collection of completely positive trace-preserving maps $\{\mathcal I_x\}_{x\in\mathcal X}$, such that the conditional state of the system $\rho_{\xi|x}$, after recording the outcome $x$, is given by $\mathcal I_x(\rho_\xi)$. Explicitly, each instrument $\mathcal I$ is described by its measurement operators $\{M_x\}_{x\in \mathcal X}$, with $M_x\in \mathsf{M}_d$ and $\Pi_x = M_x^\dagger M_x$, which determine the post-measurement state as follows, 
\be
\rho_{\xi|x}= \mathcal I_x(\rho_\xi) = \frac{M_x \rho_\xi M_x^\dagger}{\tr(\rho_\xi \Pi_x)}\;.
\ee
\par
Here, it is assumed that a single measurement operator $M_x$ corresponds to any given POVM element $\Pi_x$ and measurement outcome $x\in \mathcal X$, i.e. the measurement is \emph{fine-grained} \cite{chefles2003retrodiction}. The set of possible instruments thus corresponds to the set of measurements having the same statistics, i.e.~for which $\Pi_x = M_x^\dagger M_x$. Each instrument gives the same statistics of outcomes, but different output states after the measurement. In fact, each measurement operator $M_x$ can be written in polar form as $M_x = U_x P_x$, where $U_x$ is unitary and $P_x$ is positive semi-definite. Since $\Pi_x = M_x^\dagger M_x = P_x^2$, it follows that $P_x$ is the principal square root of $\Pi_x$, while $U_x$ is arbitrary.
\par
In the following, it is understood that any given measurement scheme $\mathscr M$ is implemented via its corresponding \emph{L\"uders instrument} $\mathcal I_{L}$, which is defined by the choice $U_x=\mathbbm I_{\mathcal H}$, $\forall x \in \mathcal X$, so that the post-measurement state is given by
\be
\rho_{\xi|x}=\mathcal I_{L,\,x}(\rho_\xi) =\frac{P_x \rho_\xi P_x}{\tr(\rho_\xi \Pi_x)}\;.
\ee
 {In other words,} since applying any instrument has the same effect as applying the L\"uders instrument, followed by a unitary control depending on the outcome of the measurement, we are restricting ourselves to \emph{bare} measurements, i.e.~with no controls.

After implementing a given measurement scheme $\mathscr M$ on $N$ identically prepared systems, the outcomes are processed through an estimator $\hat \xi$, i.e.~any measurable function $\mathcal X^{\times N} \to \Xi$. One usually focuses on the subfamily of \emph{unbiased} estimators, i.e.~estimators such that $ \text E_\xi(\hat \xi) = \xi$, $\forall \xi\in \Xi$ {, where $\text E_\xi(\cdot)$ denotes the expectation value with respect to $p_{x,\,\xi}$} \footnote{Being unbiased for all possible values $\xi\in \Xi$ is usually too strong of a condition to impose on an estimator $\hat \xi$. In some cases, in fact, no unbiased estimator exists. One can relax the condition of unbiasedness to \emph{local} unbiasedness without changing the main results that follow.}. The performance of a general estimator is quantified by choosing a loss function; its expected value  is then a measure of the estimator's performance.  A standard choice is the quadratic loss function $(\hat \xi -\xi)^2$, so that for an unbiased estimator the expected loss coincides with its variance. The best performing unbiased estimator is thus the one with minimum variance. 

The Cram\'er-Rao bound \cite{cramer2016mathematical,rao1992information} states that, under mild regularity conditions, the variance of any unbiased estimator is bounded from below by the inverse of the Fisher information $\mathcal F_\xi$ (FI), i.e. 
\be
N\cdot \text{Var}(\hat \xi) \cdot \mathcal F_\xi(\rho_\xi,\,\mathscr M)\geq 1\;,
\ee
where
\be
\mathcal F_\xi(\rho_\xi,\,\mathscr M) \eqd \text E_\xi([\partial_\xi \log p_{x,\,\xi}]^2)
\ee 
and the multiplicative factor of $N$ comes from the additivity of the FI. Any estimator achieving equality in the Cram\'er-Rao bound is called \emph{efficient}. Efficient estimators do exist, i.e.~either for finite $N$, when $p_{x,\,\xi}$ is an exponential family and $\xi$ is one of its natural parameters, or asymptotically as $N\to \infty$, e.g.~the maximum-likelihood and Bayes estimators. Moreover, there is a precise sense \cite{lecam1953some,stigler2007epic,van1998asymptotic,ibragimov2013statistical} in which, under suitable regularity conditions, in the asymptotic regime the FI sets the optimal performance of any consistent estimator, both biased and unbiased. 

It follows that generally the optimal measurement to implement is the measurement maximizing the FI. One defines the quantum Fisher information $\mathcal F_\xi^{(Q)}$ (QFI) \cite{pla95,holevo2011probabilistic,h72,braunstein1994statistical,paris09} as
\be\label{qfimax}
\mathcal F_\xi^{(Q)}(\rho_\xi) \eqd \underset{\mathscr M}{\text{max}}\;\mathcal F_\xi(\rho_\xi,\,\mathscr M)\;.
\ee
The QFI can be explicitly computed as 
\be
\mathcal F_\xi^{(Q)}(\rho_\xi)=\tr(\rho_\xi\,L_{\rho,\,\xi}^2)\;,
\ee
where $L_{\rho,\,\xi}$ is the symmetric logarithmic derivative (SLD) of $\rho_\xi$, i.e.~the Hermitian matrix satisfying the equation $\partial_\xi \rho_\xi = (\rho_\xi L_{\rho,\,\xi}+L_{\rho,\,\xi}\rho_\xi)/2$. The optimal measurement achieving the maximum in Eq.~\eqref{qfimax} is a projective measurement of the SLD.

\subsection{ {Quantifying disturbance}}\label{distdef}
 {In the following we are going to quantify disturbance by any of the following measures} $\mathfrak D_\xi^{(\alpha)}$ ($\alpha\in\{\Delta,\,F,\,\tau,\,\pi\}$).
\begin{itemize}
\item  {The $\Delta$-disturbance $\mathfrak D^{(\Delta)}_\xi$, which is the average information loss \cite{shitara2016trade}, i.e.~the measurement-induced decrease of the QFI $\mathcal F_\xi^{(Q)}$. The explicit expression is given by}
\be
\mathfrak  D^{(\Delta)}_\xi(\mathscr M) \eqd \mathcal F_\xi^{(Q)}(\rho_\xi) -
\langle \mathcal F_\xi^{(Q)}(\rho_{\xi|x}) \rangle\;,
\ee
where 
\be
 \langle \mathcal F_\xi^{(Q)}(\rho_{\xi|x}) \rangle \eqd \sum_{x\in\mathcal X} p_{x,\,\xi} \mathcal F_\xi^{(Q)}(\rho_{\xi|x})
\ee
is the QFI computed for the post-measurement state, averaged over the outcomes of the measurement.
\item   {The $F$-disturbance $\mathfrak  D^{(F)}_\xi$, corresponding to the} average fidelity-based distance \cite{fuchs1996quantum} between the initial and the post-measurement state, i.e.
\be
\mathfrak D^{(F)}_\xi(\mathscr M) \eqd 1- \sum_{x\in \mathcal X}p_{x,\,\xi} F^2(\rho_\xi,\,\rho_{\xi|x})\;,
\ee
where $F(\rho_\xi,\,\rho_{\xi|x})$ is the fidelity, 
\be
F(\rho_\xi,\,\rho_{\xi|x}) = \tr\left[\sqrt{\sqrt{\rho_\xi}\,\rho_{\xi|x}\sqrt{\rho_\xi}}\right]\;.
\ee
\item  {The $\tau$-disturbance $\mathfrak D^{(\tau)}_\beta$, which is defined for a thermal statistical model $\rho_\beta$} as the average spectral temperature variation, i.e.
\be 
\mathfrak D^{(\tau)}_\beta(\mathscr M) \eqd \sum_{x\in\mathcal X} p_{x,\,\beta}\, \abs{\beta - \tau(\rho_{\beta|x})}\;\,
\ee
where the spectral temperature $\tau(\rho)$  {of a quantum state is defined as follows} \cite{gemmer2004quantum}
\be\label{sptemp}
\begin{split}
\qquad\;\tau(\rho) \eqd & \left(1-\frac{p_{0}+p_{d-1}}{2}\right)^{-1} \\
&\quad \times \sum_{i=0}^{d-1}\left(\frac{p_{i+1}+p_{i}}{2}\right)\frac{\log(p_{i}/p_{i+1})}{E_{i+1}-E_i}\;;
\end{split}
\ee
$\{E_i\}_{i=0,\,\dots,\,d-1}$ is the energy spectrum, assumed to be non-degenerate, and $p_{i}$ is the probability of the outcome $E_i$, following a projective measurement of the Hamiltonian $H$.

\item  {The $\pi$-disturbance $\mathfrak D^{(\pi)}_\xi$, which} is the quantum relative entropy between the quantum I-projection of the post-measurement state onto the statistical model and the pre-measurement state, averaged over the outcomes of the measurement. Explicitly, 
\be
\mathfrak D^{(\pi)}_\xi(\mathscr M) \eqd \sum_{x\in\mathcal X} p_{x,\,\xi}\, D_Q (\rho_{\eta^{(\pi)}_x}||\rho_\xi) \;,
\ee
where
\be\label{etadef}
\eta_x^{(\pi)} \eqd \underset{\eta}{\text{arg min}}\, D_Q(\rho_\eta||\rho_{\xi|\,x})\;,
\ee 
and $D_Q(\cdot || \cdot)$ denotes the quantum relative entropy between two density operators.
\end{itemize}
\subsection{Typologies of quantum measurements}
In the following, a measurement scheme $\mathscr M$ is referred to as:
\begin{itemize}
\item \emph{projective}, if $\forall x\in \mathcal X$, $\Pi_x$ is a projector, i.e.~$\Pi_x^2 = \Pi_x$. It is rank-1 projective if moreover $\text{rank}(\Pi_x)=1$, $\forall x\in\mathcal X$;
\item \emph{semiclassical}, if all its POVM elements $\{\Pi_x\}_{x\in \mathcal X}$ can be simultaneously diagonalized together with the statistical model $\rho_\xi$, i.e.~$[\Pi_x,\,\rho_\xi]=[\Pi_x,\,\Pi_{x'}]=0$, $\forall x,\,x' \in \mathcal X$;
\item \emph{irreversible}, if some of its POVM elements are non-invertible matrices;
\item \emph{efficient}, with respect to a given disturbance measure 
$\mathfrak D^{(\alpha)}_\xi(\mathscr M)$, if it minimizes $\mathfrak D^{(\alpha)}_\xi(\mathscr M)$,  {for a fixed value of the extracted information} $\mathcal F_{\xi}(\mathscr M)$.  
\end{itemize}
Let us remark that, for a rank-1 projective measurement, the post-measurement state no longer depends on the parameter; as a consequence, the QFI of the post-measurement state vanishes. For a semiclassical measurement, one has $[P_x,\,\rho_\xi]=[P_x,\,P_{x'}]=0$, $\forall x,\,x' \in\mathcal X$ \footnote{This follows form the fact that, if $A,\,B\in \mathsf{Her}^+_d$ are commuting positive semidefinite matrices, then one also has $[A,\,\sqrt B]=[\sqrt A,\,\sqrt B]=0$}; as a consequence, the conditional state after a sequence of semiclassical measurements does not depend on the order they are performed.
\begin{figure}[h!]
\flushright{
\includegraphics[width=0.95\columnwidth]{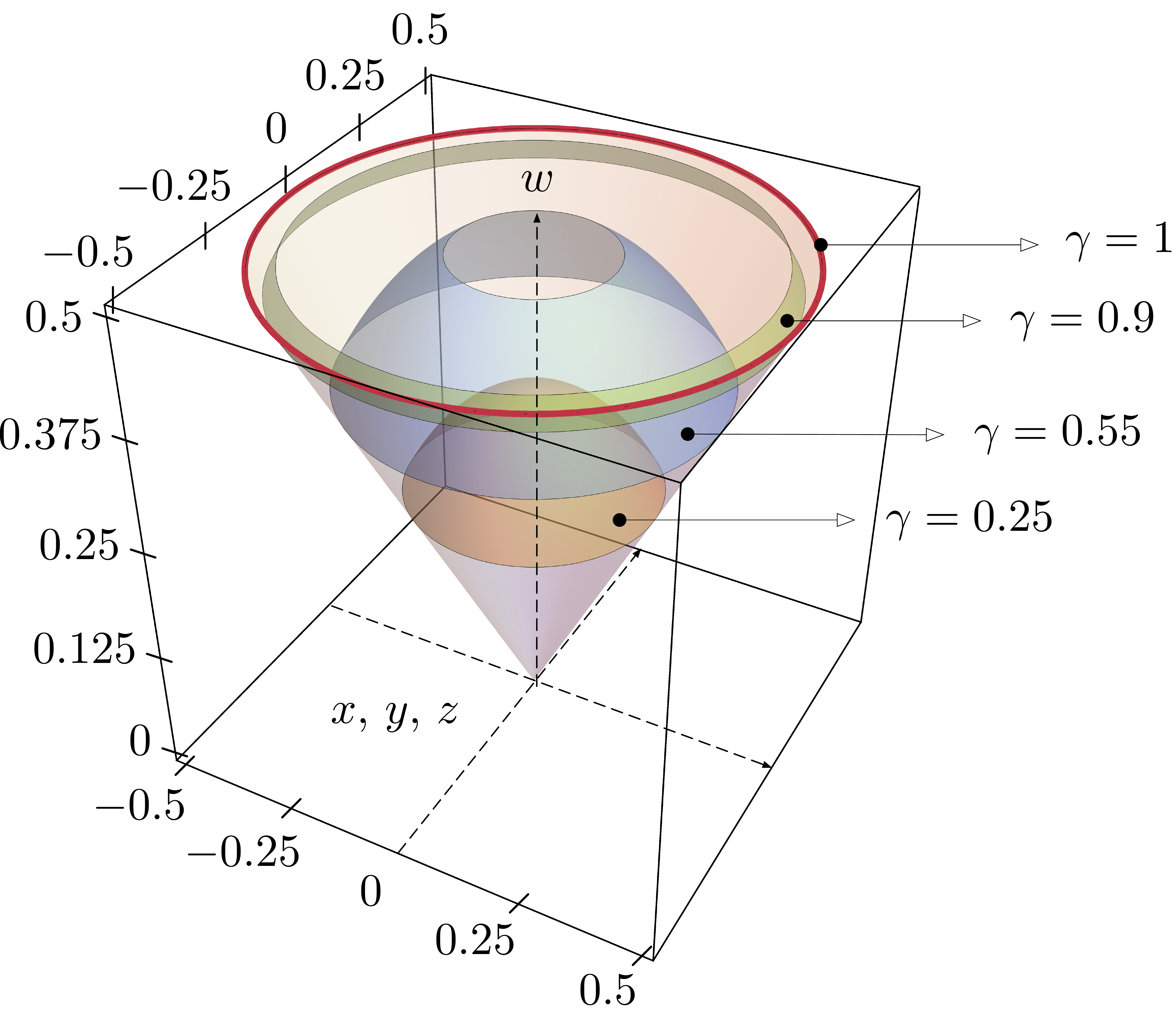}}
\caption{Graphical representation of the set of binary measurements on a qubit. Each measurement scheme $\mathscr M$ is described by a POVM $(\Pi_0,\,\Pi_1)$. Since $\Pi_1=\mathbbm I_2 -\Pi_0$, it is sufficient to specify $\Pi_0$, e.g.~via its cartesian coordinates $(w,\,x,\,y,\,z)$. Any binary POVM is thus associated with a point of the cone of equations $w^2-x^2-y^2-z^2\geq 0$ and $0< w\leq 1/2$. The spherical caps are the loci of points with fixed value of the POVM purity $\gamma$.\label{fig0}}
\end{figure}
\subsection{Qubit thermometry}
Temperature is not a quantum observable, so a parameter estimation framework is unavoidable  {to address and analyze any measurement scheme aimed at its determination.} The temperature of a thermal bath can be inferred by putting it into contact with a two-dimensional quantum system, waiting  {long enough} for it to thermalize, and then performing a suitable measurement. In the following, the ground state of the qubit system is denoted by $\ket{1}$ and the excited state by $\ket{0}$; the Hamiltonian is $H=\delta \sigma_z /2$, with $\sigma_z$ being the third Pauli matrix. The statistical model is the thermal family of equilibrium states 
\be
\rho_\beta = \frac1{Z_\beta}\text{diag}(e^{-\beta\delta/2},\,e^{\beta\delta/2})\;,
\quad Z_\beta = 2 \cosh(\beta \delta/2)\,.
\ee

Concerning the class of measurements implementable on the qubit thermometer, it is rather natural to restrict attention to binary measurements, i.e.~measurement schemes with sample space $\chi = \{0,\,1\}$. In this regard we state the following proposition.
\begin{prop}\label{binpovm}
Any binary measurement scheme on a qubit, with corresponding POVM $\{\Pi_0,\,\Pi_1\}$, is of the form 
\be\label{povmexp}
\Pi_0 = w \mathbbm I_2 + x \sigma_x + y \sigma_y +z \sigma_z\;,\qquad \Pi_1 = \mathbbm I_2 - \Pi_0\;,
\ee
where $0< w \leq 1/2$ and $\sqrt{x^2+y^2+z^2}\leq w$.
\begin{proof}
To represent a physical POVM, $\Pi_0$ and $\Pi_1$ must be positive semidefinite matrices. For $2\times 2$ matrices, this is equivalent to imposing that both their trace and determinant are nonnegative. By explicit computation, after expanding $\Pi_0$ as in Eq.~\eqref{povmexp} on the basis of $\mathsf{Her}_2$ made up of the identity matrix $\mathbbm I_2$ and the three Pauli matrices $\sigma_x,\,\sigma_y,\,\sigma_z$, one obtains the constraints $0< w \leq 1$, $\sqrt{x^2+y^2+z^2} \leq \text{min}(w,\,1-w)$. By assuming without loss of generality that $\tr(\Pi_0)\leq 1$, the previous constraints simplify to $$0< w \leq 1/2 \qquad \sqrt{x^2+y^2+z^2} \leq w\,.$$ 
\end{proof}
\end{prop}
We refer to the set $(w,x,y,z)$ as the cartesian coordinates of the corresponding POVM. It is also convenient to introduce \emph{conical} coordinates $(w,\lambda, \theta, \varphi)$ such that
\begin{align}
x &= \lambda w \sin\theta \cos\varphi\,,\notag\\
y &= \lambda w \sin\theta \sin\varphi\,,\notag\\
z &= \lambda w \cos\theta\,,
\end{align}
with the constraints $0< w\leq 1/2,\, 0\leq \lambda\leq 1,\, 0\leq \theta\leq \pi,\, 0\leq \varphi \leq 2\pi$. As a visual aid, the set of binary POVMs on a qubit can be represented as a cone in the Euclidean space $\mathbbm R^4$ with $(w,\,x,\,y,\,z)$ as cartesian coordinates (see Fig.~\ref{fig0}). Each cross-section of the cone with a hyperplane of constant $w$ is a 3-dimensional ball of radius $w$, with $\theta$ the polar angle and $\varphi$ the azimuthal angle. { Projective} measurements correspond to the sphere $w=1/2$, $\lambda=1$, whereas { irreversible} measurements to the surface of the 
cone $\lambda=1$. 
\par
A binary measurement can further be characterized in terms  {of its 
purity $\gamma$ and its non-commutativity $\chi$, defined as follows
\begin{align}
\gamma & \eqd \tr(\Pi_0^2) = 2(1+\lambda^2)w^2 \\
\chi &\eqd \sin\theta\,.
\end{align}}
The purity $\gamma$ measures the proximity to the set of projective measurements, since $\gamma=1$ precisely when $w=1/2$ and $\lambda=1$. 
The non-commutativity $\chi$ measures the distance from the set of semiclassical measurements; in fact, the POVM elements $\Pi_x$ commute with $\rho_\beta$ precisely when $\chi=0$ (or in the 
trivial case of the uninformative measurement $\lambda = 0$). Measurements that maximize the non-commutativity $\chi$, i.e.~having $\theta=\pi/2$, are referred to as \emph{non-classical}. 
\begin{figure}[h!]
\centering
\includegraphics[width=0.95\columnwidth]{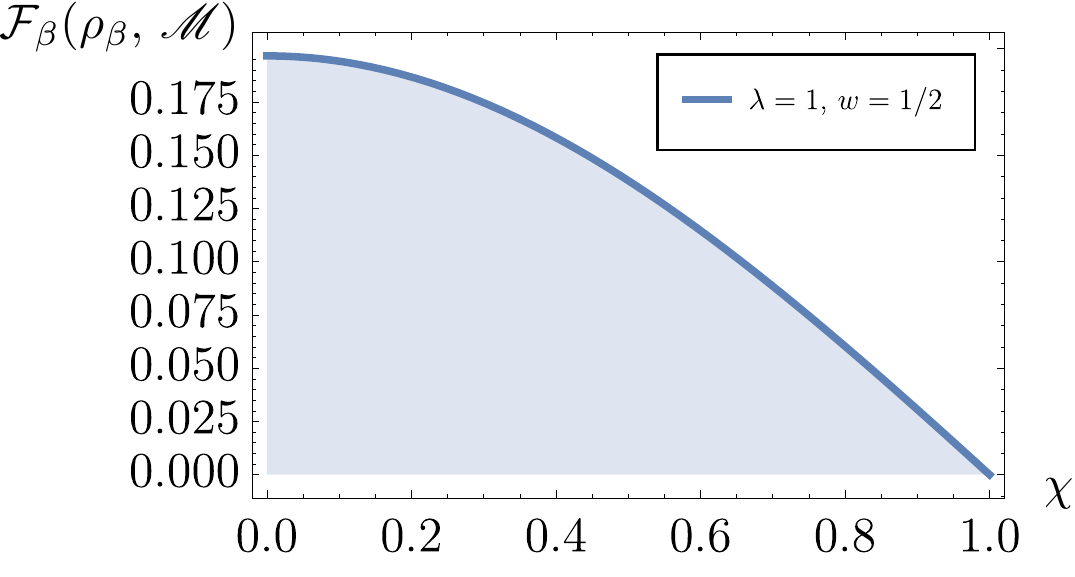}\\
\includegraphics[width=0.95\columnwidth]{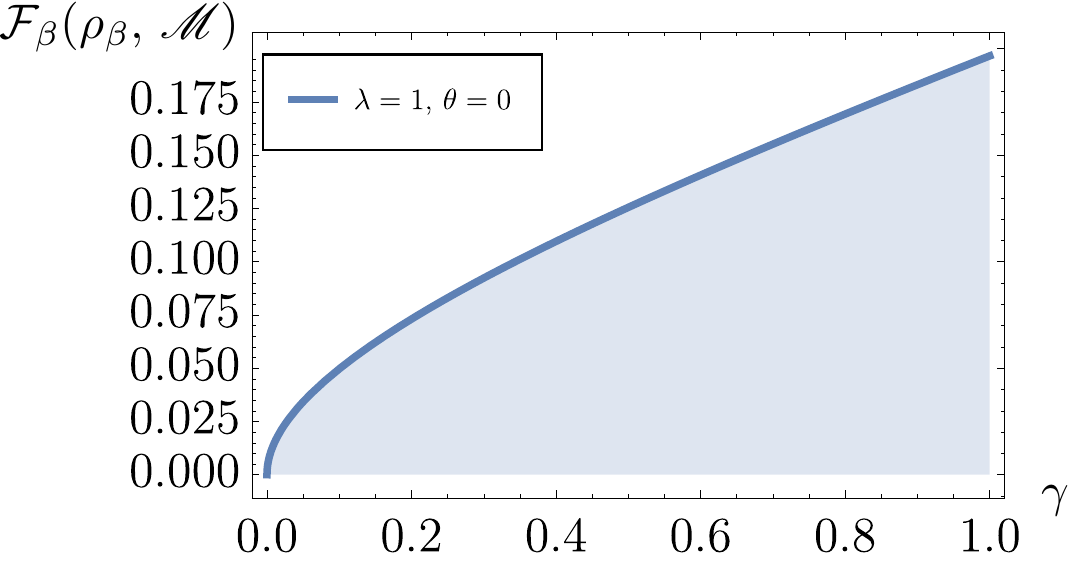}
\caption{ {\emph{Upper panel}: the Fisher information $\mathcal F_\beta$ 
as a function of the measurement non-commutativity $\chi$. 
\emph{Lower panel}: the Fisher information $\mathcal F_\beta$ 
as a function of the measurement purity $\gamma$. 
Both plots are obtained upon fixing $\beta=\delta=1$.\label{fishfig}}}
\end{figure}
\section{Qubit thermometry: information}\label{infoqt}
Given a binary measurement scheme $\mathscr M$, corresponding to conical coordinates $(w,\,\lambda,\,\theta,\,\varphi)$,  the information $\mathcal F_\beta(\rho_\beta,\,\mathscr M)$ about the temperature is given by 
\begin{align}
\label{fish}
\mathcal F_\beta(\rho_\beta,\,\mathscr M) &= 
\frac{\delta ^2\,\lambda ^2 w  \cos ^2\theta\, \text{sech}^4\left(\frac{\beta  \delta }{2}\right)}{4 \mathcal Q  \left[1-w
\mathcal Q\right]}\;, \\
\mathcal Q &=  \left[1-\lambda  \cos\theta  \tanh \left(\frac{\beta  \delta }{2}\right)\right]\;. \notag
\end{align}
In Fig.~\ref{fishfig},  {we show the range} of $\mathcal F_\beta$ as $\mathscr M$ is varied, as a function of either the non-commutativity $\chi$ or the purity $\gamma$ of the measurement. The set of all possible binary POVMs is 4-dimensional; however, the information $\mathcal F_\beta$ does not depend on the conical coordinate $\varphi$. Thus, fixing either $\chi$ or $\gamma$ leaves only two free parameters: the result is a two-dimensional region. The boundary curve of such region corresponds to measurements extracting maximum information for a given value of either $\gamma$ or $\chi$. 
\par
For instance, the measurement scheme $\mathscr F_{\chi}^{\text{max}}$ that extracts maximum information for given value of $\chi$ is obtained for $\lambda=1$ and $w=1/2$, i.e.~it is the projective POVM of the form 
\be\label{projpovm}
\Pi_0 = \frac{1}{2}\begin{pmatrix}
1\pm\sqrt{1-\chi^2}&\chi\, e^{-i\varphi}\\
\chi\, e^{i\varphi}&1\mp\sqrt{1-\chi^2}&\\
\end{pmatrix}
\;,\;\;  \Pi_1 = \mathbbm I_2-\Pi_0\;.
\ee 
Analytically, the boundary curve (shown in the upper panel of Fig. \ref{fishfig}) is given by 
\be
\mathcal F_\beta(\rho_\beta,\,\mathscr F_{\chi}^{\text{max}}) = \frac{2\beta^2\,(1-\chi^2)}{4-\chi^2+4\cosh(\beta\delta) +\chi^2 \cosh(2\beta\delta)}\;.
\ee
Similarly, the  measurement scheme $\mathscr F_{\gamma}^{\text{max}}$ that extracts maximum information for given value of $\gamma$ is obtained for $\lambda=1$ and $\theta=0$, i.e.~it is a semiclassical measurement of the form 
\be\label{semipovm}
\Pi_0 = \begin{pmatrix}
\sqrt \gamma & 0\\
0 & 0\\
\end{pmatrix} 
\;,\qquad  \Pi_1 = \mathbbm I_2-\Pi_0\;.\qquad 
\ee
Analytically, the boundary curve (shown in the lower panel of Fig. \ref{fishfig}) is given by
\be
\mathcal F_\beta(\rho_\beta,\,\mathscr F_{\gamma}^{\text{max}}) = \frac{\beta^2 \sqrt \gamma\, e^{2\beta\delta}}{(1+e^{\beta\delta})^2\,(1+e^{\beta\delta}-\sqrt\gamma)}\;.
\ee
\section{Qubit thermometry: disturbance}\label{distqt}In this section, the four disturbance measures $\mathfrak D^{(\alpha)}_\beta$, defined in Sec.~\ref{distdef} are considered and analysed in details.  {In particular, their relation to the non-commutativity and the purity of the POVM is studied, as well as their trade-off with the extracted information. We also characterize explicitly the $\alpha$-efficient classes of measurements in the four cases. }
\subsection{ {The $\Delta$-disturbance} $\mathfrak D_\beta^{(\Delta)}$}
A given measurement scheme $\mathscr M$ extracts an amount of information $\mathcal F_\beta$, which is a fraction of the total available information, i.e.~the QFI $\mathcal F_\beta^{(Q)}$. At the same time, part of the information on the parameter is lost due to the measurement, which is quantified by the disturbance $\mathfrak D_\beta^{(\Delta)}$. Using the method outlined 
in App.~\ref{appA}, it is straightforward to  {obtain that $\mathfrak D_\beta^{(\Delta)}$ is given by} 
\be\label{dist}
\mathfrak D_\beta^{(\Delta)}(\mathscr M) = 
\mathcal F_\beta^{(Q)}(\rho_\beta) - p_{0,\,\beta} \mathcal F_\beta^{(Q)}(\rho_{\beta|\,0}) - p_{1,\,\beta}\mathcal F_\beta^{(Q)}(\rho_{\beta|\,1})\;,
\ee
where  {
\begin{align}
\mathcal F_\beta^{(Q)}(\rho_\beta) & = \frac{\delta^2}{(2+2\cosh{\beta\delta)}}\,, \\ p_{0,\,\beta}& =w-\lambda w \cos\theta \tanh(\beta\delta/2)\,,\\ 
p_{1,\,\beta}& =1-p_{0,\,\beta}\;, \notag
\end{align} 
and the QFI of the conditional states are given by
\begin{align}
\mathcal F_\beta^{(Q)}(\rho_{\beta|\,0}) &= \frac{\delta^2(1-\lambda^2)\,e^{\beta\delta}}{\left(K_+ + e^{\beta\delta} K_-\right)^2}\,, \\
\mathcal F_\beta^{(Q)}(\rho_{\beta|\,1}) &=\frac{\delta^2[(1-w)^2-\lambda^2w^2]\,e^{\beta\delta}}{[1-w K_+ +e^{\beta\delta}(1-w K_-)]^2}\,, \\
K_\pm & = 1\pm \lambda\cos\theta\,. \notag
\end{align}}
The measurements that maximize $\mathfrak D_\beta^{(\Delta)}$ are the projective measurements and their {$\Delta$-disturbance} equals the QFI. The measurements that minimize $\mathfrak D_\beta^{(\Delta)}$ are the uninformative measurements ($\lambda=0$), which cause no  {$\Delta$-disturbance}.
\begin{figure}[h!]
\centering
\includegraphics[width=0.95\columnwidth]{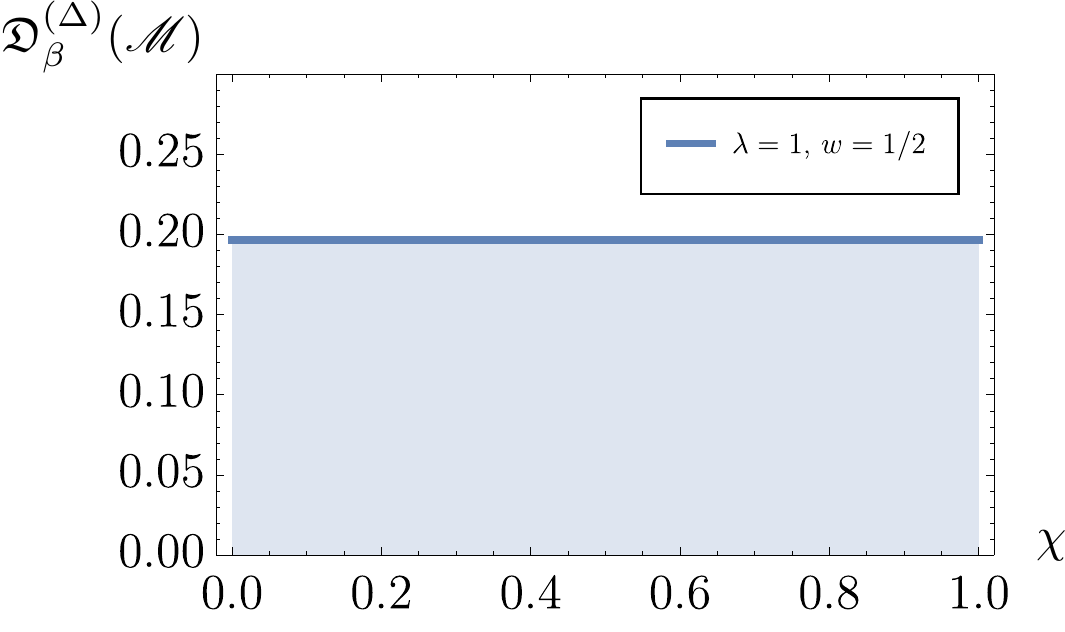}\\
\includegraphics[width=0.95\columnwidth]{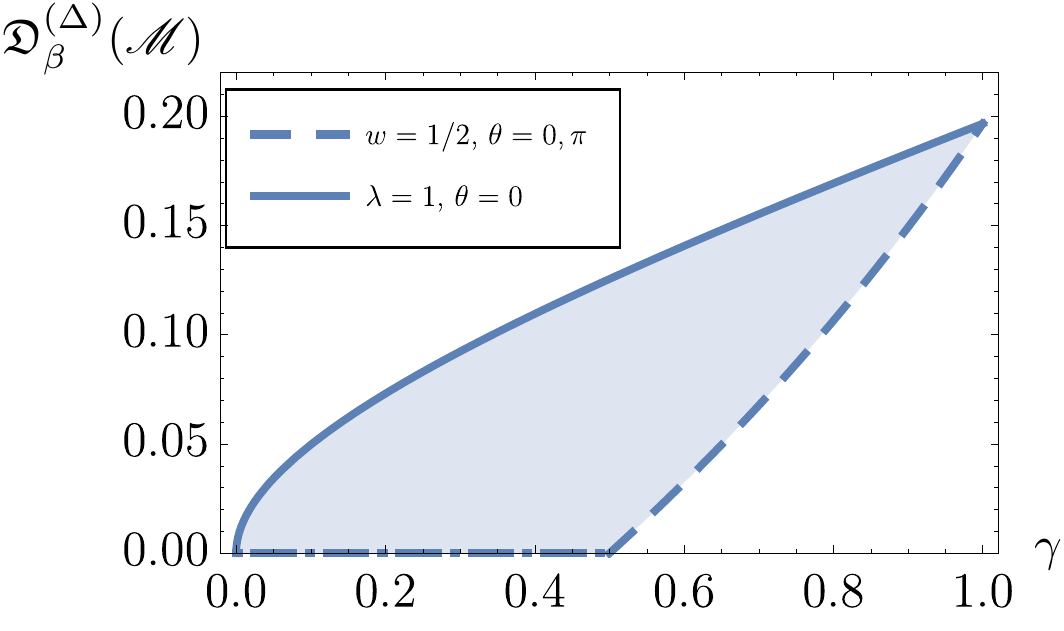}
\caption{ {\emph{Upper panel}: Range of the $\Delta$-disturbance $\mathfrak D_\beta^{(\Delta)}$ as a function of the measurement non-commutativity $\chi$. \emph{Lower panel}: a similar plot as a function of the measurement purity $\gamma$. Both plots are obtained upon fixing $\beta=\delta=1$.}\label{ddvscg}}
\end{figure}
\par
In Fig.~\ref{ddvscg},  {we show the range} of $\mathfrak D_\beta^{(\Delta)}$ while $\mathscr M$ is varied, as a function of either the non-commutativity $\chi$ or the purity $\gamma$. The  {measurement scheme $\mathscr D_{\chi}^{\text{max}}$ leading to} maximum disturbance $\mathfrak D^{(\Delta)}_\beta$ for given value of $\chi$ is obtained for $\lambda=1$ and $w=1/2$, i.e.~it is the projective POVM of Eq.~\eqref{projpovm}. Such maximum value is the QFI $\mathcal F_\beta^{(Q)}$. The scheme $\mathscr D_{\gamma}^{\text{max}}$ that causes maximum disturbance for given value of $\gamma$ is instead obtained for $\lambda=1$ and $\theta=0$, i.e.~it is the semiclassical POVM of Eq.~\eqref{semipovm}. The corresponding value of the disturbance is
\be
\mathfrak D^{(\Delta)}_\beta(\mathscr D_{\gamma}^{\text{max}}) 
= \frac{\beta^2 \sqrt\gamma\, e^{2\beta\delta}}{(1+e^{\beta\delta})^2(1+e^{\beta\delta}-\sqrt\gamma)}\;.
\ee
\par
While for any given value of $\chi$ it is possible to find a zero-disturbance measurement, measurements with purity $\gamma > 1/2$ must destroy information. For given value of $\gamma$, the minimum achievable information loss is 
\be
\mathfrak D^{(\Delta)}_\beta(\mathscr D_{\gamma}^{\text{min}}) = \frac{\beta^2\, (2\gamma-1)\; \text{sech}^2({\beta\delta}/{2})}{4 (\gamma +\cosh{\beta\delta}-\gamma \cosh{\beta\delta} )}\;.
\ee  
It is attained by measurements schemes $\mathscr D_{\gamma}^{\text{min}}$ having $w=1/2$ and $\theta=0,\,\pi$, which defines a subclass of semiclassical POVMs of the form
\begin{align}\label{povmpminfd}
\Pi_0 & = \frac{1}{2}\begin{pmatrix}
1\pm \sqrt{2\gamma-1} & 0 \\
0 & 1\mp\sqrt{2\gamma-1}
\end{pmatrix}\;, \\
\Pi_1& = \mathbbm I_2 - \Pi_0\;.
\end{align}
The choice of sign corresponds, respectively, to the case $\theta=0$ (for the upper choice of sign) or $\theta=\pi$ (for the lower choice). Let us also remark that Eq.~\eqref{povmpminfd} actually describes a unique physical POVM, since a relabelling of the outcomes $0\to 1$ (and $1\to 0$) interchanges the two POVMs for $\theta=0,\,\pi$.
\begin{figure}[h!]
\includegraphics[width=1.00\columnwidth]{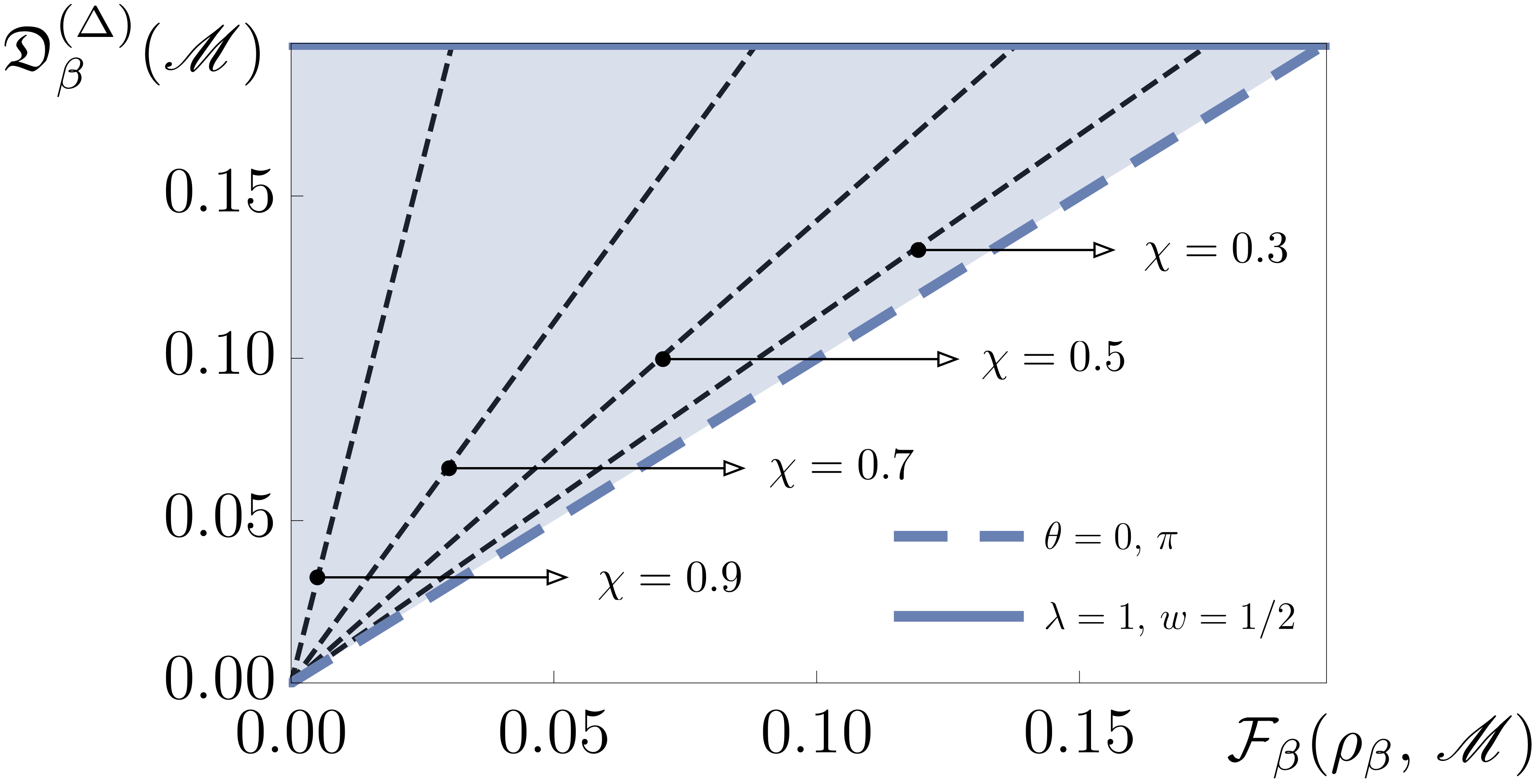}
\includegraphics[width=0.78\columnwidth]{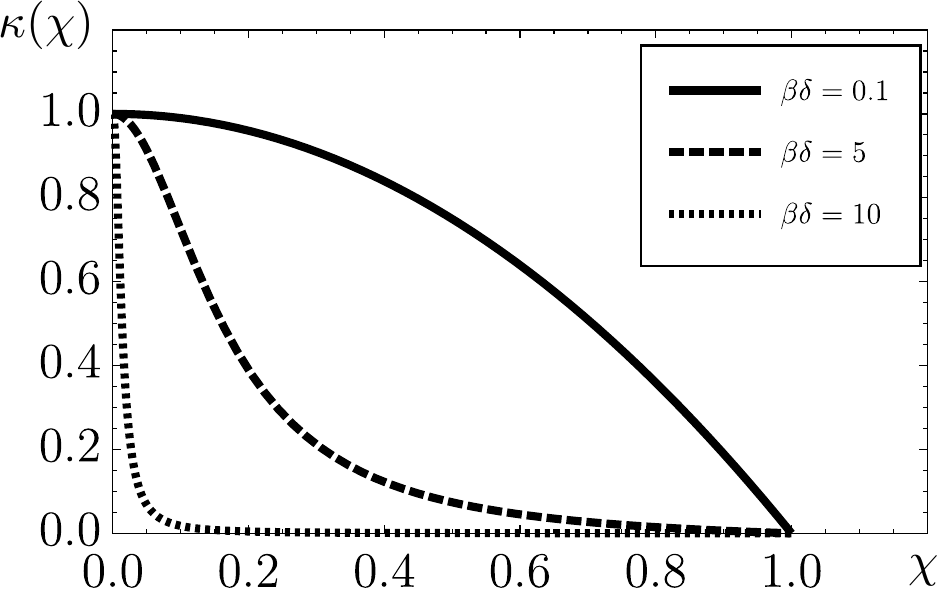}
\caption{ {\emph{Upper panel}: information/disturbance trade-off region for $\mathfrak D_\beta^{(\Delta)}$. The region is foliated into line segments by fixing the value of the non-commutativity $\chi$. \emph{Lower panel}: the quantity $\kappa(\chi)$ (see Proposition \ref{tor} for details) for different values of the product $\beta \delta$.}\label{ddto}}
\end{figure}
\subsubsection{ {The information/$\Delta$-disturbance trade-off -- $\mathcal F_\beta$ vs $\mathfrak D_\beta^{(\Delta)}$}}
It is often the case that one is not interested in $\mathcal F_\beta$ or $\mathfrak D_\beta^{(\Delta)}$, taken individually, but rather in their trade-off. Since both the information $\mathcal F_\beta$ and the information loss $\mathfrak D_\beta^{(\Delta)}$ are independent from $\varphi$, the trade-off region in the plane $\mathcal F_\beta$ vs $\mathfrak D_\beta^{(\Delta)}$ corresponds to 3 free parameters. However, it can be foliated into a set of 1-dimensional curves by fixing one additional parameter, which turns out to be the non-commutativity $\chi$.  {The situation is summarized by  the following proposition.}
\begin{prop}
\label{tor}
The trade-off region in the plane $\mathcal F_\beta$ vs $\mathfrak D^{(\Delta)}_\beta$ is the triangle of vertices $(0,\,0)$, $(0,\,\mathcal F_\beta^{(Q)})$ and $(\mathcal F_\beta^{(Q)},\,\mathcal F_\beta^{(Q)})$. For fixed value of the non-commutativity $\chi$, the resulting trade-off curve is a line segment with endpoints $(0,\,0)$ and $(\kappa(\chi)\,\mathcal F_\beta^{(Q)},\,\mathcal F_\beta^{(Q)})$, where
\be\label{kappa}
\kappa(\chi) = \frac{2(1-\chi^2)}{2-\chi^2+\chi^2\cosh(\beta\delta)}\;.
\ee
\begin{proof}
By direct computation, from Eqs.~\eqref{fish} and \eqref{dist}, one finds that the ratio between $\mathcal F_\beta(\rho_\beta,\,\mathscr M)$ and $\mathfrak D_\beta^{(\Delta)}(\mathscr M)$ is a function only of $\chi$, denoted by $\kappa(\chi)$, where $\kappa(\chi)$ is given in Eq.~\eqref{kappa}. Since, for any fixed value of $\chi$, the $\Delta$-disturbance $\mathfrak D_\beta^{(\Delta)}$ ranges from $0$ to $\mathcal F_\beta^{(Q)}$, it follows that the trade-off region is foliated into line segments with endpoints $(0,\,0)$ and $(\kappa(\chi)\,\mathcal F_\beta^{(Q)},\,\mathcal F_\beta^{(Q)})$. Since the range of $\kappa(\chi)$ is the interval $[0,\,1]$, it follows that the trade-off region, which is the union of the trade-off curves for fixed $\chi$, is the triangle of vertices $(0,\,0)$, $(0,\,\mathcal F_\beta^{(Q)})$ and $(\mathcal F_\beta^{(Q)},\,\mathcal F_\beta^{(Q)})$ (see Fig.~\ref{ddto}). 
\end{proof}
\end{prop}
From Prop.~\ref{tor}, it follows in particular that the disturbance of an efficient measurement equals the corresponding extracted information. In Ref.~\cite{shitara2016trade}, it was proven in full generality that, for a parameter $\xi$ and statistical model $\rho_\xi$, $\mathcal F_\xi$ and $\mathfrak D_\xi^{(\Delta)}$ satisfy the inequality $\mathcal F_\xi(\rho_\xi,\,\mathscr M) \leq \mathfrak D_\xi^{(\Delta)}(\mathscr M)$. In the specific case of qubit thermometry, Prop.~\ref{tor} implies that such inequality is tight. In fact, saturation occurs when $\kappa(\chi)=1$, or equivalently $\chi=0$, i.e.~the measurement is semiclassical. It therefore follows that the set of efficient measurements coincides with the set of semiclassical measurements. In fact, it holds more generally, for any $d$-dimensional thermometer, that semiclassical measurements are efficient according to the disturbance measure $\mathfrak D^{(\Delta)}_\beta$. 
\begin{prop}
For a thermal statistical model of a $d$-level system, semiclassical measurements are efficient according to the disturbance measure $\mathfrak D^{(\Delta)}_\beta$.  
\begin{proof}
The statistical model is $\rho_\beta = e^{-\beta H}/\tr(e^{-\beta H})$, where $H\in \mathsf{Her}_d$ is the Hamiltonian of the system. Its SLD is $L_{\rho,\,\beta}= \langle H\rangle_\beta- H$, where $\langle H\rangle_\beta = \tr(H \rho_\beta)$; thus, in the eigenbasis of $H$, both $\rho_\beta$ and $L_{\rho,\,\beta}$ are diagonal matrices. Now, consider a semiclassical measurement scheme $\mathscr M$ with POVM $\{\Pi_x\}_{x\in\mathcal X}$ and post-measurement state $\rho_{\beta|x} = \rho_\beta \Pi_x/p_x$, where $P_x=\sqrt{\Pi_x}$. Its derivative is $\partial_\beta \rho_{\beta|x} = (L_{\rho,\,\beta}-\partial_\beta \log p_x)\rho_{\beta|x}$. Since each $\Pi_x$ commutes by definition with $\rho_\beta$, it must commute also with $H$, thus with $L_{\rho,\,\beta}$. It follows that $L_{\rho,\,\beta}$ also commutes with $\rho_{\beta|x}$, and so the SLD of the conditional state $\rho_{\beta|x}$ is $L_{\rho,\,\beta|x} = L_{\rho,\,\beta}-\partial_x \log p_x \mathbb I_d$. The average QFI $\langle \mathcal F_\xi^{(Q)}(\rho_{\xi|x}) \rangle$ of the post-measurement state can now be expanded as
\be
\begin{split}
\langle \mathcal F_\xi^{(Q)}(\rho_{\xi|x}) \rangle=&
\sum_{x\in\mathcal X}\tr(\rho_\beta\,	 L_{\rho,\,\beta}^2\Pi_x) +
\sum_{x\in\mathcal X} (\partial_\beta \log p_x)^2 p_x\\
&\quad -2 \sum_{x\in\mathcal X} (\partial_\beta \log p_x) \tr(L_{\rho,\,\beta}\,\rho_\beta \Pi_x)\;.
\end{split}
\ee
The first sum is equal to the QFI $\mathcal F_\beta^{(Q)}(\rho_\beta)$. The second sum is equal to the FI $\mathcal F_\beta(\rho_\beta,\,\mathscr M)$. The third term is also equal to the FI, after making use of the fact that $\tr(L_{\rho,\,\beta}\,\rho_\beta \Pi_x)=\partial_\beta p_x$. It follows immediately that the information $\mathcal F_\beta$ saturates to the disturbance $\mathfrak D_\beta^{(\Delta)}$, and the measurement is efficient. 
\end{proof}
\end{prop}
\subsection{ {The $F$-disturbance} $\mathfrak D_\beta^{(F)}$}
\begin{figure}[h!]
\centering
\includegraphics[width=0.95\columnwidth]{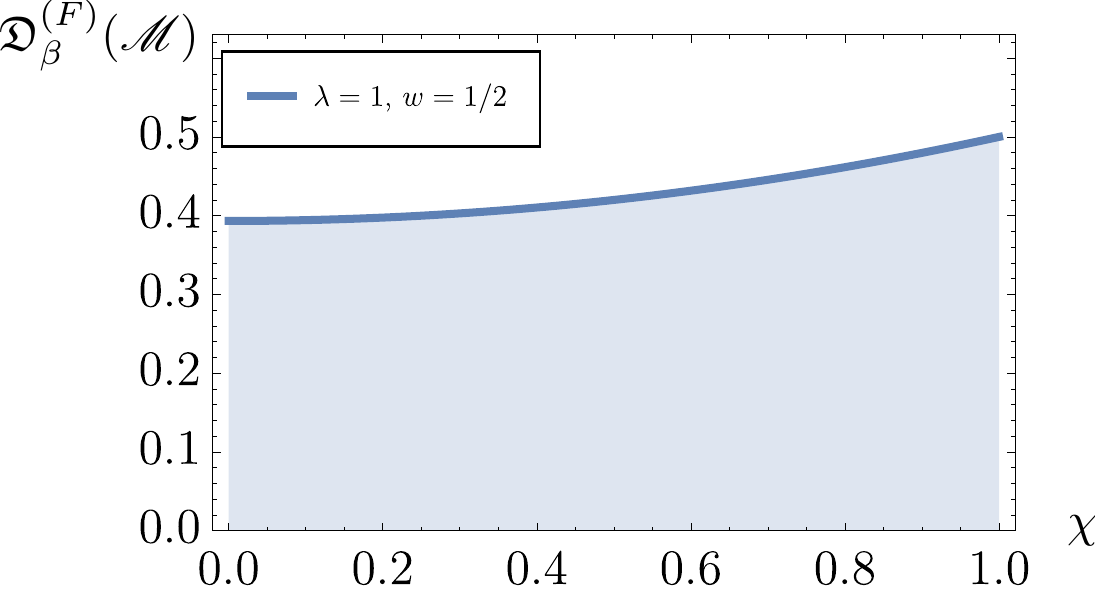}\\
\includegraphics[width=0.95\columnwidth]{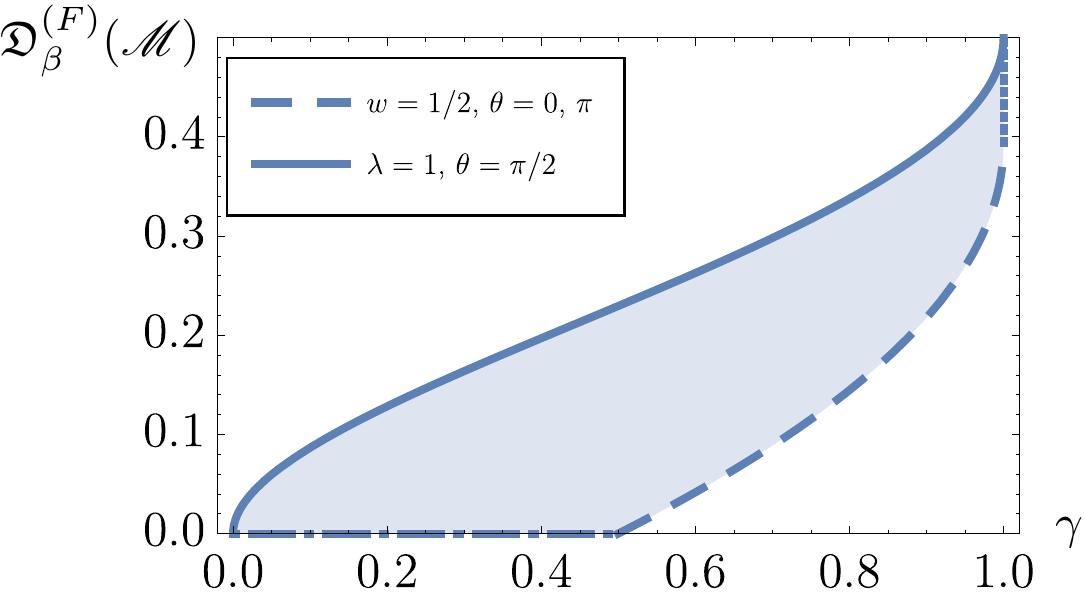}
\caption{\label{fdist}
 {Range of the $F$-disturbance $\mathfrak D_\beta^{(F)}$ as a function of the measurement non-commutativity $\chi$ (upper panel) or the measurement purity $\gamma$ (lower panel). Both plots are obtained upon fixing $\beta\delta=1$.}}
\end{figure}

The $F$-disturbance $\mathfrak D_\beta^{(F)}$ is defined as the average (fidelity-based) distance between the statistical model and the post-measurement state. Explicitly, in our model, it is given by
\be
\begin{split}
\mathfrak D_\beta^{(F)}(\mathscr M) = &\frac{1+\cos^2\theta+\sin^2\theta\cosh(\beta\delta)}{2[1+\cosh(\beta\delta)]} \\
&\times \left[1-w\sqrt{1-\lambda^2}-\sqrt{(1-w)^2-w^2\lambda^2}\right]\;.
\end{split}
\ee
The measurements that maximize $\mathfrak D_\beta^{(F)}$ are the 
projective measurements which have $\chi = 1$, corresponding to 
POVMs of the form
\be
\Pi_0 = \frac{1}{2} \begin{pmatrix}
1 & e^{-i\varphi}\\
e^{i \varphi} & 1\\
\end{pmatrix}\;,\qquad \Pi_1 = \mathbbm I_2 -\Pi_0\;.
\ee
The maximum $F$-disturbance is thus $\mathfrak D_\beta^{(F)}=1/2$. The measurements that minimize $\mathfrak D_\beta^{(F)}$ are the uninformative measurements, which cause no $F$-disturbance. If the non-commutativity $\chi$ is fixed, the maximum achievable $F$-disturbance is attained by a projective POVM and is equal to
\be
\mathfrak D_\beta^{(F)}(\mathscr D^{\text{max}}_\chi) = \frac{2-\chi^2+\chi^2\cosh(\beta\delta)}{2+2\cosh(\beta\delta)}\;.
\ee
If instead the purity $\gamma$ is fixed, the maximum achievable $F$-disturbance is attained by POVMs with $\lambda=1$ and $\theta=\pi/2$, i.e.~of the form	
\be
\Pi_0 = \frac{\sqrt\gamma}{2} \begin{pmatrix}
1 & e^{-i\varphi}\\
e^{i \varphi} & 1\\
\end{pmatrix}\;,\qquad \Pi_1 = \mathbbm I_2 -\Pi_0\;.
\ee
The corresponding disturbance is
\be
\mathfrak D_\beta^{(F)}(\mathscr D_\gamma^{\text{max}}) = \frac{1-\sqrt{1-\sqrt\gamma}}{2}\;.
\ee
For $\gamma>1/2$, there is also a non-trivial lower bound to the $F$-disturbance,
\be
\mathfrak D_\beta^{(F)}(\mathscr D_\gamma^{\text{min}}) = \frac{1-\sqrt{2-2\gamma}}{1+\cosh(\beta\delta)}\;,
\ee
which is achieved by a special subclass of semiclassical measurement schemes having $w=1/2$ (and $\theta=0,\,\pi$), corresponding to the POVMs already given in Eq.~\eqref{povmpminfd}. Thus, for fixed purity, the least disturbing measurements are the same, for both disturbance measures $\mathfrak D_\beta^{(\Delta)}$ and $\mathfrak D_\beta^{(F)}$.
{The range of $\mathfrak D_\beta^{(F)}$ while $\mathscr M$ is varied, is illustrated in Fig.~\ref{fdist}, as a function of either the non-commutativity $\chi$ or the purity $\gamma$ of the POVM.}
\subsubsection{ {The information/$F$-disturbance trade-off -- $\mathcal F_\beta$ vs $\mathfrak D_\beta^{(F)}$}}
\begin{figure}[h!]
\flushleft{
\includegraphics[width=0.95\columnwidth]{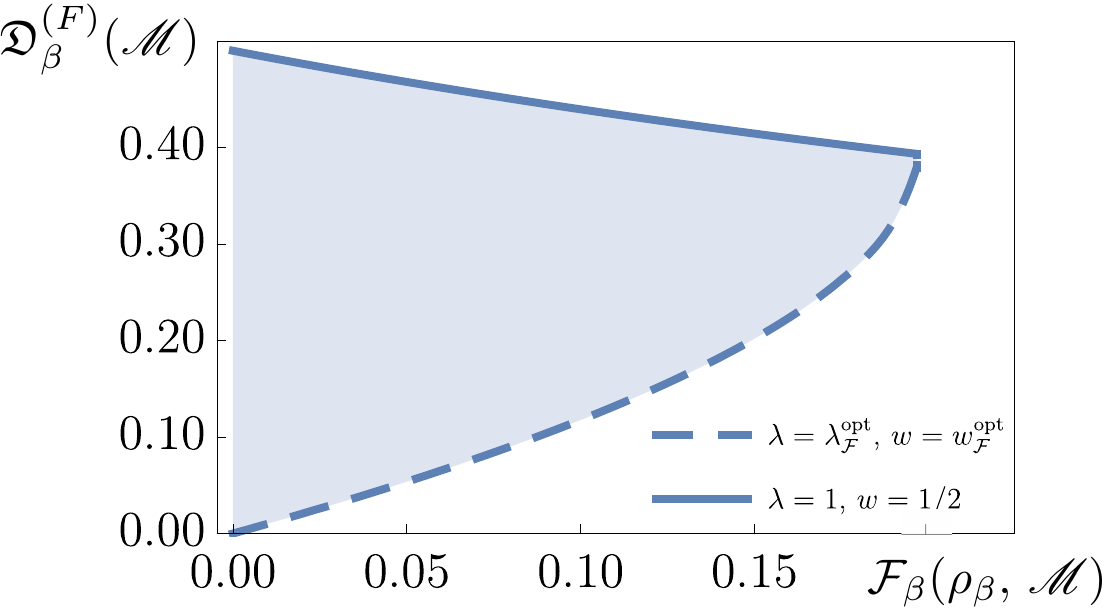}}
\flushright{
\includegraphics[width=0.9\columnwidth]{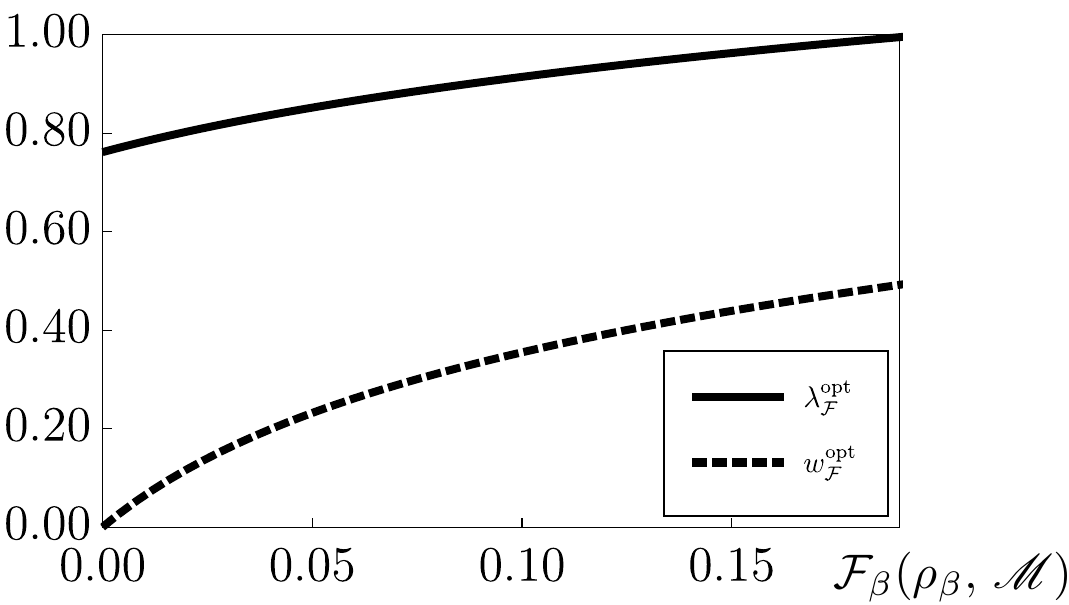}}
\caption{ {\emph{Upper panel}: trade-off region for the $F$-disturbance 
measure $\mathfrak D_\beta^{(F)}$. The $F$-efficient measurements (\emph{dashed}) correspond to $\theta=0$, while $\lambda=\lambda^{\text{opt}}_{\mathcal F}$ and $w=w^{\text{opt}}_{\mathcal F}$ are determined numerically. \emph{Lower panel}: parameters $\lambda^{opt}_\mathcal F$ and $w^{opt}_\mathcal F$ of the $F$-efficient POVMs as a function of the extracted information.\label{dfto}}}
\end{figure}

The trade-off region for $\mathfrak D_\beta^{(F)}$ is not as simple to describe as it has been for $\mathfrak D_\beta^{(\Delta)}$ in Prop.~\ref{tor}; however, it is qualitatively similar. The $F$-efficient measurements are a subset of semiclassical measurements: they correspond to POVMs having $\theta=0$, while the optimal values of $\lambda$ and $w$ (denoted by $\lambda^{\text{opt}}_{\mathcal F}$ and $w^{opt}_\mathcal F$) must be determined numerically. The trade-off is illustrated in Fig.~\ref{dfto}. The lower curve represents $F$-efficient measurement schemes and has been reconstructed by minimizing numerically the $F$-disturbance, for fixed value of the extracted information, and then interpolating between the points thus obtained. The upper curve is made up of the projective measurements, which maximize the $F$-disturbance for given information. Explicitly, the maximum $F$-disturbance for fixed information $\mathcal F_\beta$ is given by
\be
\mathfrak D_\beta^{(F)}(\mathscr D^{\text{max}}_{\mathcal F}) = \frac{\beta^2\delta^2}{2\beta^2\delta^2-\mathcal F_\beta +\mathcal F_\beta \cosh(2\beta\delta)}\;.
\ee 
In the same figure, the behaviors of $\lambda^{\text{opt}}_{\mathcal F}$ and $w^{opt}_\mathcal F$ for $F$-efficient measurements, as a function of $\mathcal F_\beta$, are also shown.
\subsection{ {The $\tau$-disturbance $\mathfrak D_\beta^{(\tau)}$}} In quantum thermometry, the post-measurement state usually does not belong to the family of thermal states, i.e.~the measurement forces the state out of equilibrium. Therefore, it cannot be assigned a temperature in the conventional sense. For non-equilibrium quantum systems, the spectral temperature, defined in Eq.~\eqref{sptemp}, is a candidate generalization to the standard temperature of equilibrium thermodynamics. The spectral temperature coincides with the standard temperature when evaluated on equilibrium states and it shares many of its thermodynamical properties \cite{gemmer2004quantum}. The $\tau$-disturbance $\mathfrak D_\beta^{(\tau)}$ is equal to the average spectral temperature variation due to the measurement. Explicitly, in our model, we have
\be
\mathfrak D_\beta^{(\tau)}(\mathscr M) = p_{0,\,\beta}\abs{\beta- \tau(\rho_{\beta|0})}+p_{1,\,\beta}\abs{\beta- \tau(\rho_{\beta|1})}\;, 
\ee
where $p_{0,\,\beta}=w-\lambda w \cos\theta \tanh(\beta\delta/2)$ , $p_{1,\,\beta}=1-p_{0,\,\beta}$, and the spectral temperatures of the two conditional states are given by
 {\begin{align}
\tau(\rho_{\beta|0})&=\frac{1}{\delta}\,\log\left[\frac{2(1-\Lambda_+\Lambda_-)+e^{\beta\delta}\Lambda_-^\theta}{2(1-\Lambda_+\Lambda_-)\,e^{\beta\delta}+\Lambda_+^\theta}\right]\,, \\
\Lambda_\pm &= \sqrt{1\pm\lambda}\,, \notag \\
\Lambda_\pm^\theta &= \Big[\Lambda_\pm\,(\cot\theta+\csc\theta)+\Lambda_\mp\,\tan(\theta/2)\Big]^2\,, \notag
\end{align}}
and
 {\begin{align}
\tau(\rho_{\beta|1})&=\frac{1}{\delta}\,\log\left[\frac{2(1-w-W_+W_-)+e^{\beta\delta}W_+^\theta}{2(1-w-W_+W_-)\,e^{\beta\delta}+W_-^\theta}\right]\,, \\
W_\pm &= \sqrt{1-w\pm \lambda w}\,, \notag \\
W_\pm^\theta &= \Big[W_\pm\,(\cot\theta+\csc\theta)+W_\mp\,\tan(\theta/2)\Big]^2 \notag\,.
\end{align}}
\par
\begin{figure}[h!]
\centering
\includegraphics[width=0.95\columnwidth]{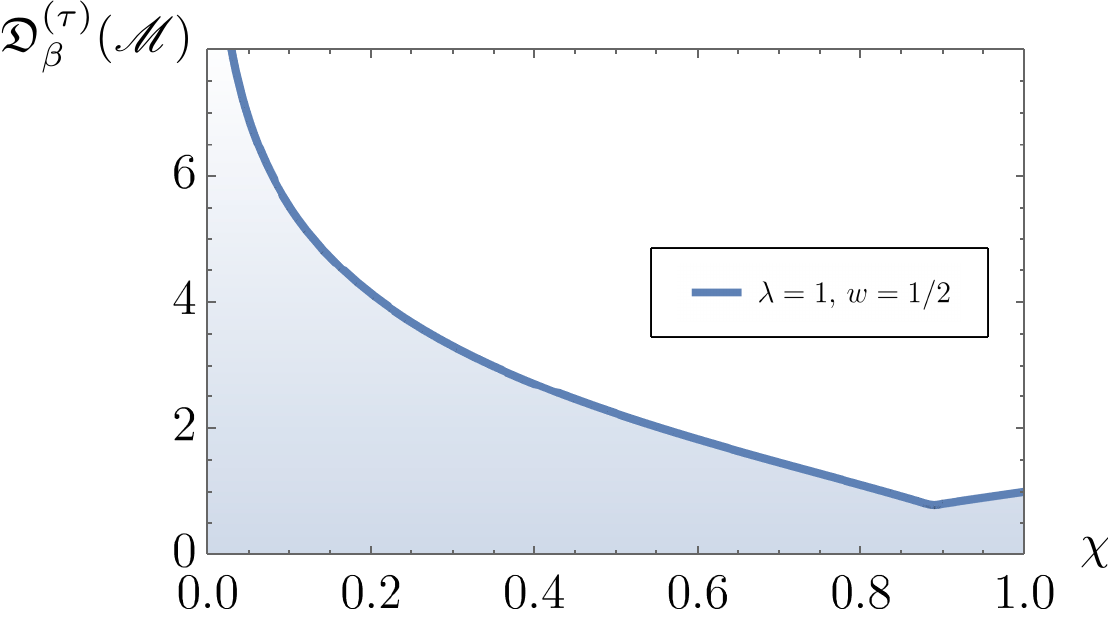}\\
\includegraphics[width=0.95\columnwidth]{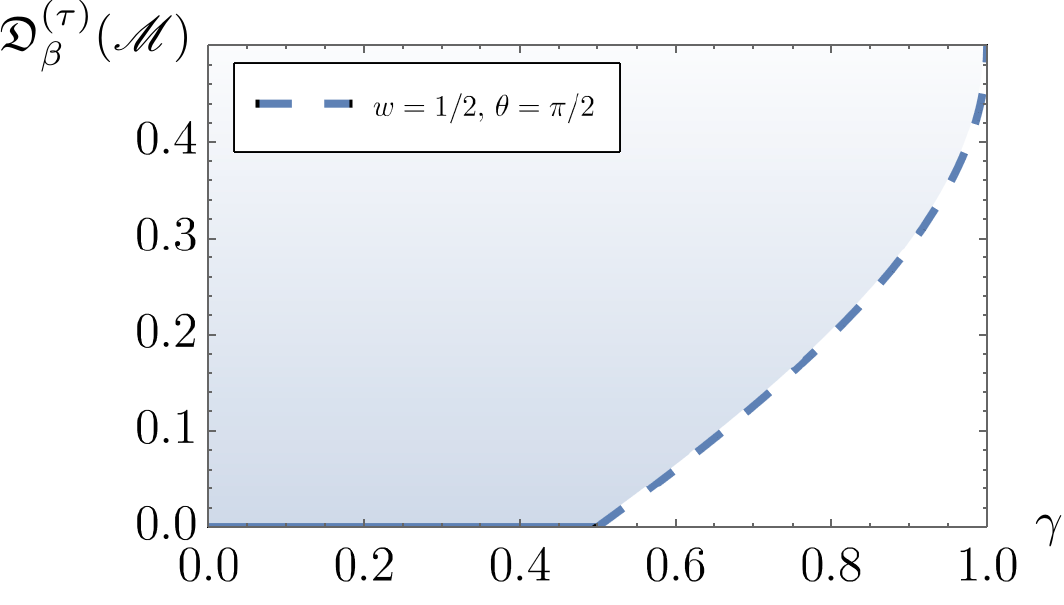}
\caption{Range of the $\tau$-disturbance $\mathfrak D_\beta^{(\tau)}$ as a function of the measurement non-commutativity $\chi$ (upper panel) or the measurement purity $\gamma$ (lower panel). Both regions are unbounded.\label{dtvscp}}
\end{figure}

In Fig.~\ref{dtvscp}, the range of $\mathfrak D_\beta^{(\tau)}$ is plotted as either the non-commutativity $\chi$ or the purity $\gamma$ are varied. For given value of $\chi$, the measurements that introduce maximum disturbance are the projective measurements; the corresponding disturbance $\mathfrak D_\beta^{(\tau)}(\mathscr D^{\text{max}}_{\chi})$  {is given by
\begin{align}
\mathfrak D_\beta^{(\tau)}(\mathscr D^{\text{max}}_{\chi}) & = 
\frac{\beta}{2} \left( H_- \widetilde{H}_- + H_+ \widetilde{H}_+ \right)
\,, \\
H_\pm & = \abs{1 \pm \frac{1}{\beta\delta}\log\frac{1-\sqrt{1-\chi^2}}{1+\sqrt{1-\chi^2}}} \notag \\
\widetilde{H}_\pm & = \left(1\pm\sqrt{1-\chi^2} \tanh\frac{\beta\delta}{2}\right)\notag
\end{align}}
Notice that there is a vertical asymptote for $\chi=0$, i.e.~there exist semiclassical measurements that cause infinite disturbance. For given $\gamma$, instead, there is no upper-bound. For $\gamma>1/2$, there is however a non-trivial lower bound. It is given by measurements that maximize the non-commutativity: their POVMs have $w=1/2$ and $\theta=\pi/2$, or explicitly
\begin{align}
\Pi_0 & = \frac{1}{2}\begin{pmatrix}  
1 & -\sqrt{2\gamma-1}\, e^{-i\varphi}\\
-\sqrt{2\gamma-1}\, e^{i\varphi} & 1
\end{pmatrix}\,, \\  \Pi_1 &= \mathbbm I_2 - \Pi_0\;, \notag
\end{align}
 {with corresponding disturbance given by
\begin{align}
\mathfrak D_\beta^{(\tau)}(\mathscr D^{\text{min}}_\gamma) & = \frac{1}{2\delta}\abs{\beta \delta -\log \frac{L_+}{L_-}}\;, \\ 
L_\pm & = 1\pm \sqrt{2-2\gamma}\,\tanh\left(\frac{\beta\delta}{2}\right)\;. \notag
\end{align}}
\subsubsection{ {The information/$\tau$-disturbance trade-off -- $\mathcal F_\beta$ vs $\mathfrak D_\beta^{(\tau)}$}}
\begin{figure}[h!]
\includegraphics[width=0.99\columnwidth]{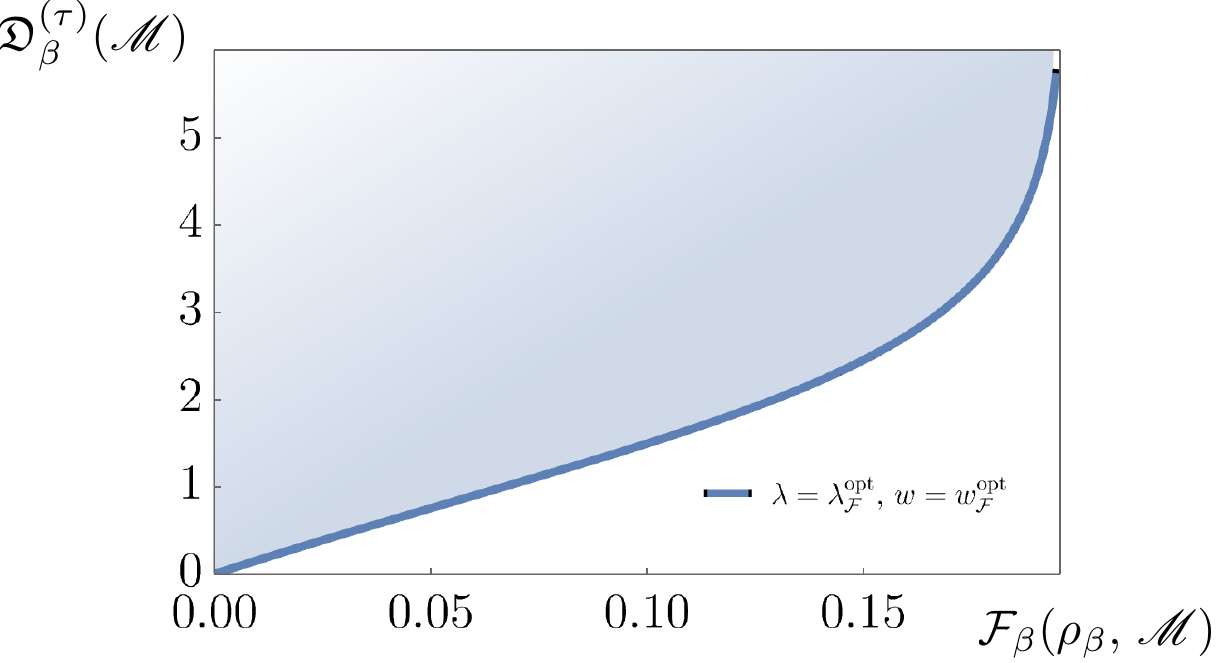}\\
\flushright{\includegraphics[width=0.86\columnwidth]{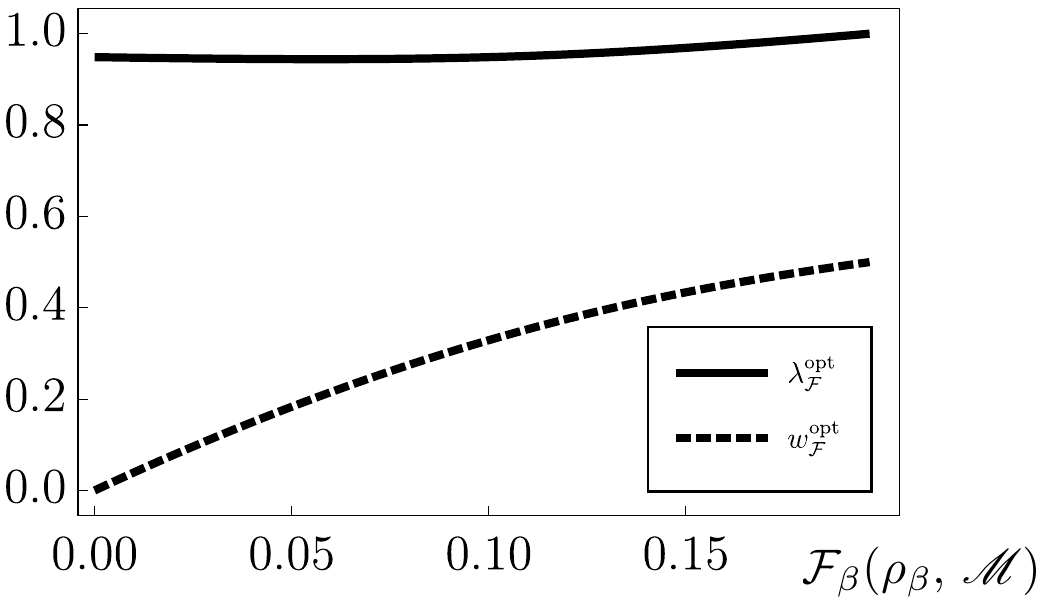}}
\caption{\emph{Upper panel}: the trade-off region for the $
\tau$-disturbance measure $\mathfrak D_\beta^{(\tau)}$. The $\tau$-efficient measurements correspond to $\theta=0$, while $\lambda=\lambda^{\text{opt}}_{\mathcal F}$ and $w=w^{\text{opt}}_{\mathcal F}$ are determined numerically. \emph{Lower panel}: behavior of the optimal parameters $\lambda^{\text{opt}}_{\mathcal F}$ and $w^{\text{opt}}_{\mathcal F}$ as a function of the FI $\mathcal F_\beta$ for $\tau$-efficient measurements. \label{dtto}} 
\end{figure}

First of all, it is easy to construct, for any given value of the extracted information, a suitable measurement with divergent $\tau$-disturbance $\mathfrak D_\beta^{(\tau)}$. Consider, e.g.~the irreversible and semiclassical measurement having $\lambda=1$,  $\theta=0$ and $w$ such that the resulting information is equal to a fixed value $\mathcal F_\beta$. Such a value of $w$ always exists and explicitly it is given by 
\be
w = \frac{(1+e^{\beta\delta})^3\mathcal F_\beta}{2[(1+e^{\beta\delta})^2\mathcal F_\beta+\beta^2\delta^2\,e^{2\beta\delta}]}\;.
\ee 
It may be easily checked that, for such POVM, the probability $p_{0,\,\beta}$ of finding the post-measurement state $\rho_{\beta|0}$ in the ground state vanishes, so that $\tau(\rho_{\beta|0})=-\infty$, and thus the $\tau$-disturbance diverges. Incidentally, such a measurement scheme is the same found in Eq.~\eqref{semipovm}, i.e.~the measurement maximizing the information for given value of the purity. Thus, there is no upper-curve to the information/$\tau$-disturbance trade-off region. {The $\tau$-efficient measurements, which minimize the disturbance for given extracted information, are a subset of the semiclassical ones.} They have $\theta=0$ and parameters $\lambda^{\text{opt}}_{\mathcal F}$ and $w^{\text{opt}}_{\mathcal F}$ which are determined numerically. The situation is summarized in Fig.~\ref{dtto}.  
\subsection{ {The $\pi$-disturbance $\mathfrak D_\beta^{(\pi)}$}}
The $\pi$-disturbance 
$\mathfrak D_\beta^{(\pi)}$ has an information-geometrical interpretation, which we briefly comment upon. The statistical distinguishability between any two equilibrium states lying on the manifold of thermal states $\rho_\beta$, is quantified via their quantum relative entropy,
\be
\begin{split}
D_Q(\rho_\eta||\rho_\beta) &= \tr(\rho_\eta \log \rho_\eta) - \tr(\rho_\eta \log \rho_\beta)\\
& = (\beta-\eta)\langle H \rangle_\eta + \log\left(Z_\beta/Z_\eta\right)\;.
\end{split}
\ee
The post-measurement state, however, is out of equilibrium. It must be projected back onto the manifold of thermal states, according to the natural geometry defined by $D_Q$. The 
$\pi$-disturbance $\mathfrak D_\beta^{(\pi)}$ is then given by the quantum relative entropy between the projected state and the original thermal state, averaged over the outcomes of the measurement. Computation of $\mathfrak D^{(\pi)}_\beta(\mathscr M)$ must in general be performed numerically \cite{niekamp2013computing}.
\par
The behavior of $\mathfrak D_\beta^{(\pi)}$ is quantitatively different, though the overall picture is qualitatively similar to the other disturbance metrics considered before. {In particular, the measurements introducing minimum $\pi$-disturbance depend on the parameter being kept fixed, however the efficient measurements are always of the semiclassical type. }

\begin{figure}[h!]
\flushright{\includegraphics[width=0.92\columnwidth]{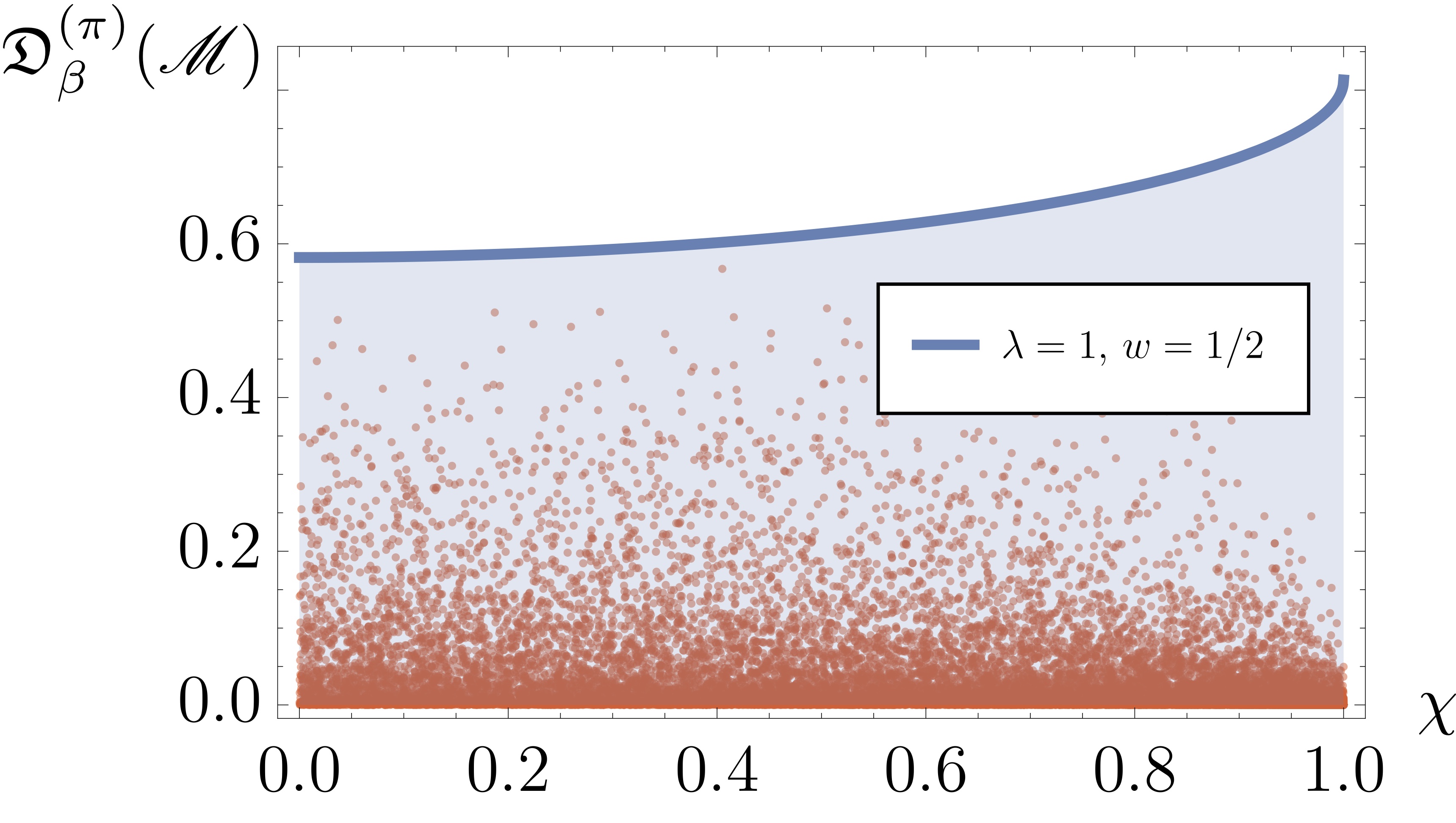}}
\includegraphics[width=0.97\columnwidth]{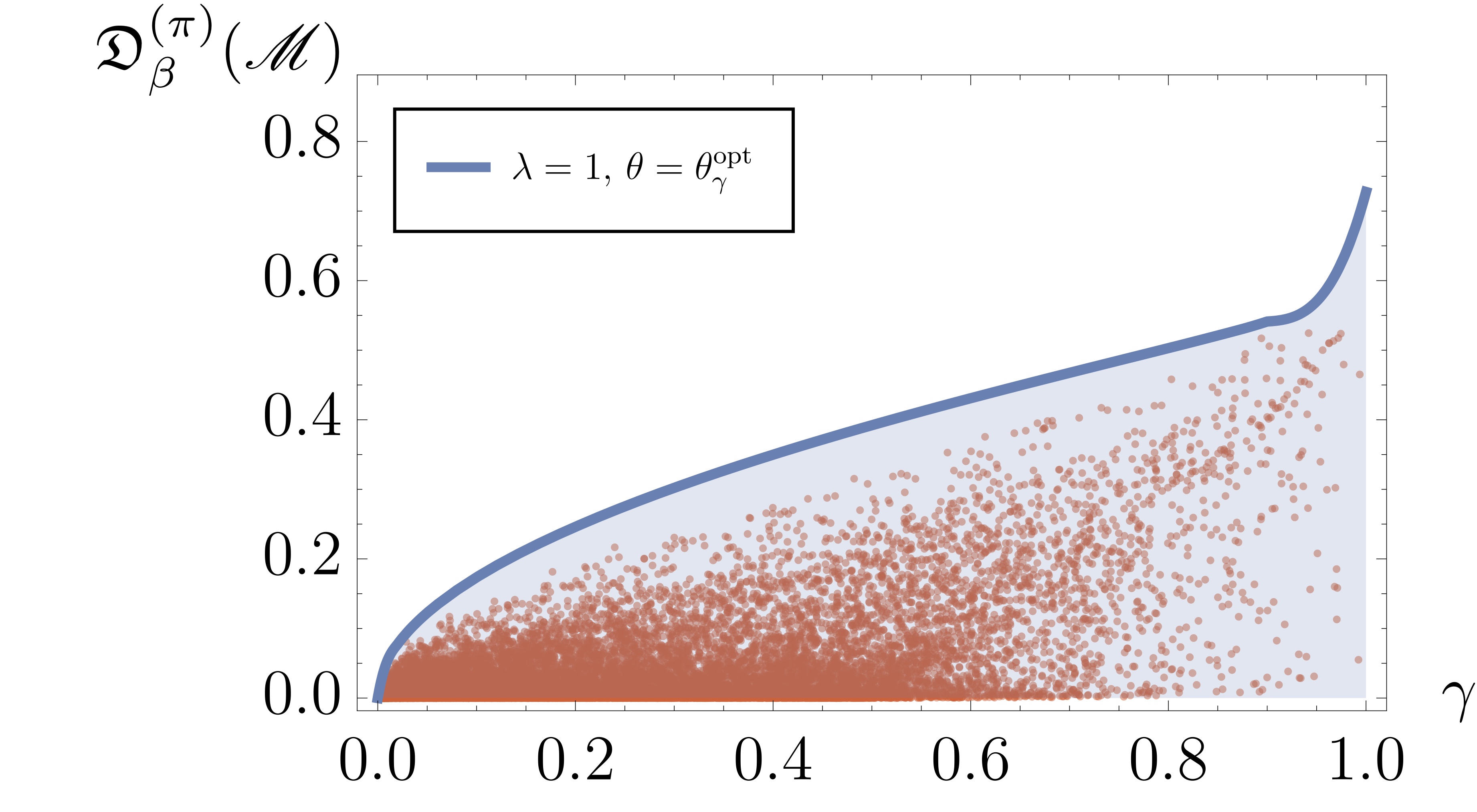}
\caption{Range of the $\pi$-disturbance
$\mathfrak D_\beta^{(\pi)}$ as either $\chi$ or 
$\gamma$ are varied. Dots correspond  {to performances of randomly 
generated POVMs with parameters $(\lambda,\,\theta,\,w)$ chosen 
uniformly.}\label{divscp}}
\end{figure}

Let us summarize the main features. For fixed value of the non-commutativity $\chi$, the measurements that maximize the disturbance are the projective ones. The corresponding disturbance  {can be computed analytically,
\begin{align}
\mathfrak D_\beta^{(\pi)}(\mathscr D^{\text{max}}_{\chi}) &=
\frac{1}{2}\left[P^\theta_- P_+ + P^\theta_+ P_-\right]\,, \\
P_\pm & = \log(1+e^{\pm\beta\delta})\,, \notag \\
P_\pm^\theta & = 
1\pm \cos\theta\, \tanh\left(\frac{\beta\delta}{2}\right) \notag\,.
\end{align}}
For fixed value of the purity $\gamma$, the measurements that maximize the disturbance are instead a subset of the irreversible ones, i.e.~they correspond to POVMs having $\lambda=1$ and $\theta=\theta_\gamma^{\text{opt}}$ determined numerically. It is also worth remarking that, contrary to the three disturbance measures previously considered, there is no nontrivial lower bound to the disturbance for $\gamma > 1/2$. In fact, the disturbance $\mathfrak D_\beta^{(\pi)}$ can be made to vanish by implementing a non-classical measurement scheme ($\theta=\pi/2$). Such measurements however do not extract any nonzero information, since the FI of Eq.~\eqref{fish} also vanishes for $\theta=\pi/2$. The situation is summarized in Fig.~\ref{divscp}.
\subsubsection{ {The information/$\pi$-disturbance trade-off -- $\mathcal F_\beta$ vs $\mathfrak D_\beta^{(\pi)}$}} 
\begin{figure}[h!]
\includegraphics[width=0.98\columnwidth]{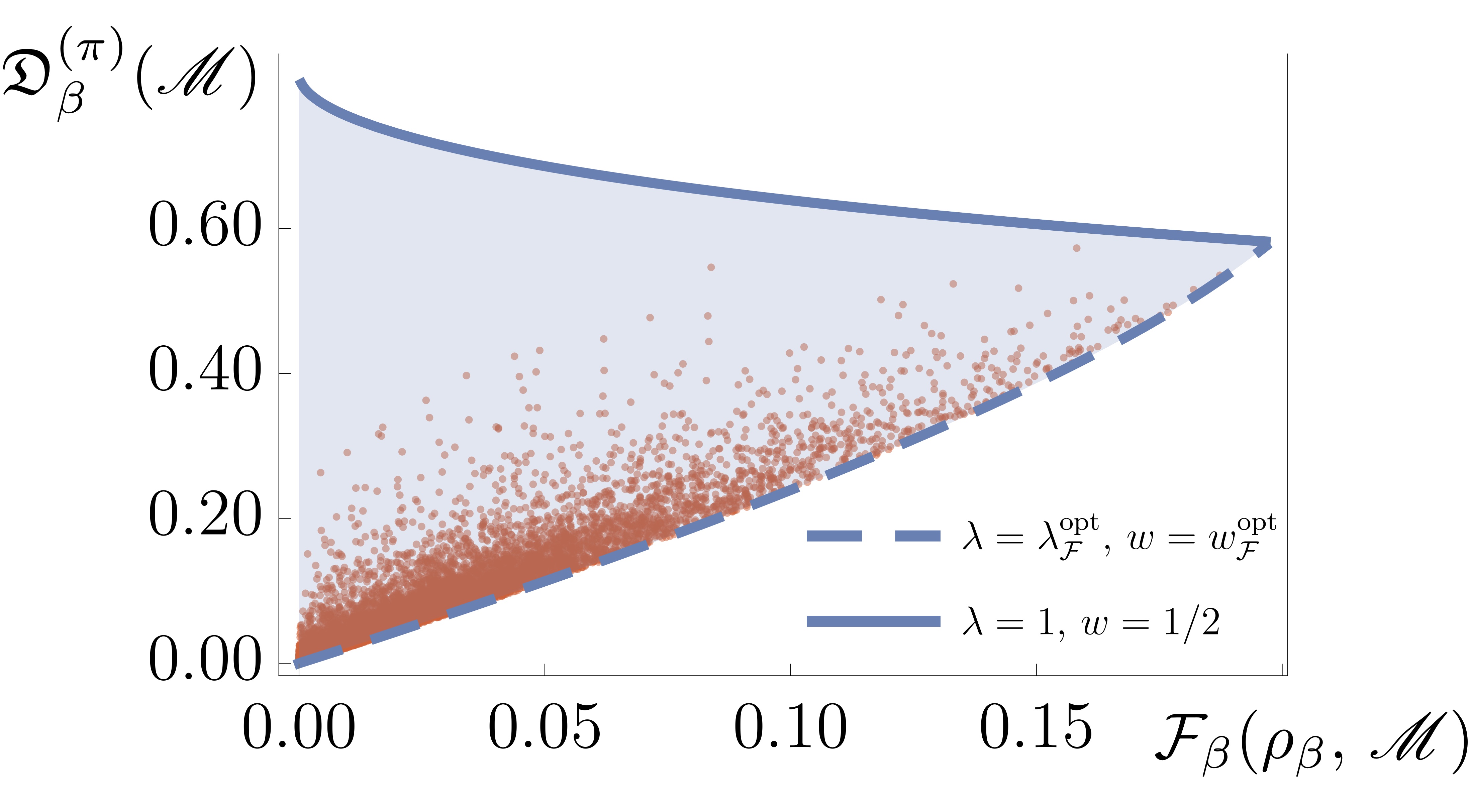}\\
\flushright{\includegraphics[width=0.9\columnwidth]{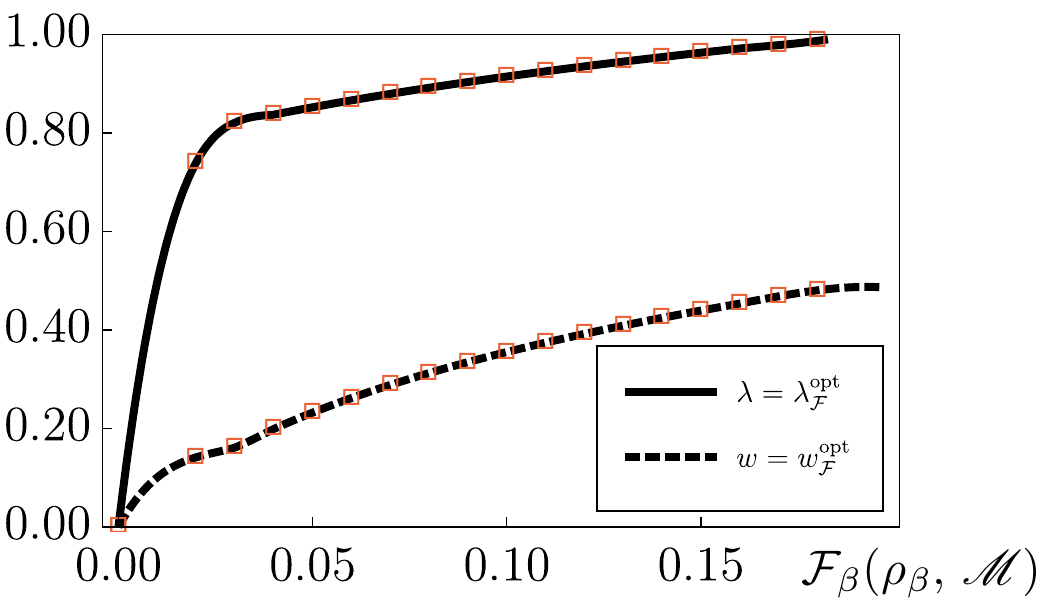}}
\caption{ {\emph{Upper panel}: trade-off region for the 
$\pi$-disturbance measure $\mathfrak D_\beta^{(\pi)}$. Dots correspond to the performances of random measurements, whose POVMs have been generated 
with uniformly randomized parameters $(\lambda,\,\theta,\,w)$. 
The $\pi$-efficient measurements (\emph{dashed}) correspond to $\theta=0$, while $\lambda=\lambda^{\text{opt}}_{\mathcal F}$ and $w=w^{\text{opt}}_{\mathcal F}$ are determined numerically. \emph{Bottom}: behavior of the optimal parameters $\lambda^{\text{opt}}_{\mathcal F}$ and $w^{\text{opt}}_{\mathcal F}$ as a function of the FI $\mathcal F_\beta$ for the $\pi$-efficient 
measurements.\label{dito}} }
\end{figure}

Concerning the information/$\pi$-disturbance trade-off, the $\pi$-efficient measurements are a subset of the semiclassical measurements. They correspond to POVMs having $\theta=0$, while the values of $\lambda$ and $w$ are found numerically. The trade-off region is shown in Fig.~\ref{dito}, together with the behaviors of the optimal parameters $\lambda^{\text{opt}}_{\mathcal F}$ and $w^{\text{opt}}_{\mathcal F}$ as a function of the FI.
\section{Conclusions}\label{concl}
In this paper, we have addressed the trade-off relation between the information on an unknown parameter, extracted via quantum measurements, and the disturbance that the probing system suffers as a result. In particular, we have analyzed in details the specific model of qubit thermometry, as a natural scenario where such trade-off plays out. 
\par
It is worth recalling at this point the main assumptions of our analysis. We have employed a two-level quantum system and performed a read-out of the thermometer via a suitable measurement. The set of measurements considered is made up of measurements which are binary (each measurement has two possible outcomes), fine-grained (there is one measurement operator for each measurement outcome) and bare (no feedback control is allowed). Such assumptions are suggested by the nature of the problem and by considerations of simplicity. In particular, while the latter two assumptions have no effect on the computation of the FI extracted by a given POVM, they allow for a greater mathematical control in the discussion of the corresponding disturbance.

Four different disturbance measures have been introduced and discussed. Our results have shown that they capture different, but consistent aspects of the trade-off relation.  {A measurement is efficient if it causes a disturbance not greater than any other measurement extracting the same amount of information. The families of efficient measurements for qubit thermometry, with respect to all four disturbance measures, have been explicitly determined. They represent different subsets of the family of semiclassical measurements, i.e.~POVMs commuting with the pre-measurement equilibrium state of the thermometer. Table \ref{tabr} summarizes results regarding the families of measurements which either minimize or maximize the disturbance, according to the four different measures defined in Subsec.~\ref{distdef}. Each measurement is denoted by $\mathfrak D_\bullet^{*}$, where $*$ stands for either minimum or maximum, while $\bullet$ is the parameter kept fixed, i.e.~either the non-commutativity $\chi$, the purity $\gamma$ or the extracted information $\mathcal F_\beta$. }
\par
Commutativity with the statistical model appears to be a necessary, but not sufficient condition for efficiency. However, it becomes sufficient in the special case when the disturbance is quantified by the information-loss $\mathfrak D^{(\Delta)}_\beta$, arguably  the most natural measure from a parameter estimation perspective. We leave it as an open question for the future whether semiclassical measurements are efficient with respect to $\mathfrak D^{(\Delta)}_\beta$ in more general scenarios, e.g. for higher dimensional thermometers or when the measurement is performed before thermal equilibrium sets in. 
\par {Our results provide novel insight on the fundamental problem of quantifying the trade-off between information and disturbance and pave the way for modeling efficient quantum thermometers, tailored to different needs.}
\begin{table*}[ht!]
 {
\caption{Summary of results about measurements minimizing or maximizing the disturbance, according to the four different measures considered in this paper. Each measurement is denoted by $\mathfrak D_\bullet^{*}$, where $*$ stands for either minimum or maximum, while $\bullet$ is the parameter kept fixed, i.e.~either the non-commutativity $\chi$, the purity $\gamma$ or the extracted information $\mathcal F_\beta$.The table specifies the typology of the measurement, i.e.~uninformative (each POVM element is proportional to the identity matrix), projective, semiclassical (each POVM element commutes with the statistical model $\rho_\beta$), non-classical (maximizes the non-commutativity $\chi$) or irreversible (the POVM elements are non-invertible matrices). More details are found directly in the main text.
\label{tabr}}
\bigskip
\centering\setlength\tabcolsep{2pt}
\begin{tabular}{|c|c|c|c|c|c|c|c|}
& $\mathscr D^{\text{min}}_{\chi}$ & $\mathscr D^{\text{max}}_{\chi} $& $\mathscr D^{\text{min}}_{\gamma < 1/2}$ & $\mathscr D^{\text{min}}_{\gamma > 1/2}$ & $\mathscr D^{\text{max}}_{\gamma}$ & $\mathscr D^{\text{min}}_{\mathcal F}$ & $\mathscr D^{\text{max}}_{\mathcal F}$\\
\hline
$\mathfrak D_\beta^{(\Delta)}$ & uninformative & projective & uninformative & $\subset$ semiclassical & $\subset$ irreversible &  semiclassical  & projective\\
$\mathfrak D_\beta^{(F)}$ &  uninformative & projective & uninformative & $\subset$ semiclassical & $\subset$  irreversible & $\subset$ semiclassical & projective \\
$\mathfrak D_\beta^{(\tau)}$ &  uninformative & projective & uninformative & $\subset$ non-classical & $\subset$  irreversible & $\subset$ semiclassical & $\subset$  irreversible\\
$\mathfrak D_\beta^{(\pi)}$ &  uninformative & projective & uninformative & $\subset$ non-classical & $\subset$ irreversible & $\subset$ semiclassical & projective\\
\hline
\end{tabular}
}
\end{table*}
\begin{acknowledgments}
This work has been supported by EU through the collaborative Project QuProCS (Grant Agreement 641277)  {and by JSPS through project 
S17118 "Geometric foundation of quantum estimation". 
MGAP is member of GNFM-INdAM. We thank Alessandro Dalle Sasse, Chiara Macchiavello and Massimo Palma for discussions in the early stage of this work.}
\end{acknowledgments}
\bibliography{refs_idtqt}
\appendix
\section{QFI for a two-level system}\label{appA}
In this appendix, an explicit expression for the QFI of a two-level system is derived. After introducing the relabellings $\sigma_0=\mathbbm I_2$, $\sigma_1 = \sigma_x$, $\sigma_2=\sigma_y$ and $\sigma_3=\sigma_z$, the statistical model $\rho_\xi$ is expanded on the basis of $\mathsf{Her}_2$ made up of the matrices $\{\sigma_\mu\}_{\mu\in\{0,1,2,3\}}$ as $\rho_\xi = \rho_\xi^{(\mu)} \sigma_\mu$., where a repeated Greek index always implies a summation on it. Similarly, its SLD $L_{\rho,\,\xi}$ is rewritten as $ L_{\rho,\,\xi} = L_{\rho,\,\xi}^{(\mu)}\, \sigma_\mu$. Let us remark that, since $\tr\rho_\xi=1$, then $\rho_{\xi}^{(0)} =1/2$, while the remaining Bloch components can be computed via $\rho_\xi^{(i)}=\tr(\rho_\xi\sigma_i)/2$. Recalling the defining relation of $L_{\rho,\,\xi}$ and employing the trace identity $\tr(\sigma_\mu\sigma_\nu)=2\delta_{\mu\nu}$, one finds that
\be\label{sldexp}
\partial_\xi \rho_\mu =  \frac{1}{2}\Re \tr\left(L_{\rho,\,\xi}\,\rho_\xi\,\sigma_\mu\right)\;.
\ee
Using the fact that $\tr(\sigma_\mu\sigma_\nu\sigma_\lambda)=2i\epsilon_{0\mu\nu\lambda}+2\delta_{\{\mu\nu}\delta_{\lambda\}0}-4\delta_{\mu 0}\delta_{\nu 0}\delta_{\lambda 0}$ (where a summation over even permutations of the indices enclosed in braces is understood), one can rewrite Eq.~\eqref{sldexp} as
\be\label{meq}
\partial_\xi \rho_\mu = M_{\mu\nu}L_{\rho,\,\xi}^{(\nu)}\;,\qquad \text{with} \qquad M_{\mu\nu} = \rho_{\{\mu}\delta_{\nu 0\}} - \delta_{\mu 0} \delta_{\nu 0}\;. 
\ee
Assuming that $M$ is invertible, which is the case when the statistical model $\rho_\xi$ has purity strictly less than 1, the Bloch components of $L_{\rho,\,\xi}$ can be computed by matrix inversion from Eq.~\eqref{meq}, i.e.~$L_{\rho,\,\xi}^{(\mu)}= (M^{-1})_{\mu\nu}\,\partial_\xi \rho_\xi^{(\nu)}$.

In turn, the Bloch components of $L_{\rho,\,\xi}$ are all that is needed to compute the QFI, since
\be\label{qfi2ls}
\mathcal F_\xi^{(Q)}(\rho_\xi)= \tr\left(\rho_\xi L_{\rho,\,\xi}^2\right)= -\big[L_{\rho,\,\xi}^{(0)}\big]^2+ \sum_{i=1}^3\,\big[L_{\rho,\,\xi}^{(i)}\big]^2\;.
\ee
To derive Eq.~\eqref{qfi2ls}, it is necessary to use the fact that $0=\tr(\rho_\xi L_{\rho,\,\xi})= 2\,\rho_\xi^{(\mu)} L_{\rho,\,\xi}^{(\mu)}$. 

Finally, substituting the explicit expression for $L^{(\mu)}_{\rho,\,\xi}$ obtained by inverting Eq.~\eqref{meq} back in Eq.~\eqref{qfi2ls}, one finds
\begin{widetext}
\be\label{qfi2lsfin}
\mathcal F_\xi^{(Q)}(\rho_\xi) = \frac{1}{4}\,\frac{\sum_{i=1}^3\big(\partial_\xi \rho_\xi^{(i)}\big)^2+4\sum_{i\neq j}\big(\partial_\xi \rho_\xi^{(i)} \partial_\xi \rho_\xi^{(j)} \rho_\xi^{(i)} \rho_\xi^{(j)} -\partial_\xi \rho_\xi^{(i)} \rho_\xi^{(j)}\big)}{1-4\sum_{i=1}^3\big(\rho_\xi^{(i)}\big)^2}\;.
\ee 
Eq.~\eqref{qfi2lsfin} allows to compute directly the QFI of a two-dimensional statistical model $\rho_\xi$, with no need to diagonalize it.
\end{widetext}
\end{document}